%% file: A-IEEETrans.tex
\documentclass[10pt, journal, twocolumn]{IEEEtran}
\input{C-macro.tex}

\usepackage{xcolor}
\usepackage{orcidlink}
\usepackage{booktabs}       
\usepackage{array}          
\usepackage{subfigure}
\usepackage{subcaption}
\usepackage[ruled]{algorithm2e}
\usepackage{algorithmic}
\usepackage[table]{xcolor}
\usepackage{color}
\usepackage{enumitem}
\usepackage{pdflscape}

\usepackage{amsthm,amsmath,amssymb}
\usepackage{adjustbox}
\usepackage{color}
\usepackage{soul}
\definecolor{mygreen}{RGB}{217,234,255} 
\sethlcolor{mygreen} 
\renewcommand{\alg}{E-MIA}

\definecolor{HLgreen}{RGB}{220,255,220}

\begin{document}

\title{E-MIA: Exam-Style Black-Box Membership Inference Attacks against RAG Systems}

\author{
  Zelin~Guan, 
  Shengda~Zhuo, 
  Zeyan~Li,
  Jinchun~He,
  Wangjie~Qiu,
  Zhiming~Zheng,
  Shuqiang~Huang
  \thanks{
    Zelin~Guan, Shengda Zhuo, and Shuqiang Huang are with the College of Cyber Security, 
    Jinan University, Guangzhou, China. 
    E-mail: a508954254@gmail.com, zhuosd96@gmail.com, hsq@jnu.edu.cn
  } 
  \thanks{
    Zeyan Li is with the School of Computer Science, 
    Shanghai Jiaotong University, Shanghai, China. 
    E-mail: zeyan0823@gmail.com
  }
  \thanks{
  Jinchun He, Wnagjie Qiu, Qinnan Zhang, and Zhiming Zheng are with the Institute of Artificial Intelligence, Beijing Advanced Innovation Center for Future Blockchain and Privacy Computing, Beihang University, Beijing 100191, China, also with Zhongguancun Laboratory, Beijing, China (e-mail: hejinchun@buaa.edu.cn; wangjieqiu@buaa.edu.cn; zzheng@pku.edu.cn.
  *(Zelin Guan and Shengda Zhuo contributed equally to this work.) 
  \emph{(Corresponding author: Shengda Zhuo and Shuqinag Huang)}
  }
}



\maketitle

\begin{abstract}
    Retrieval-Augmented Generation (RAG) equips large language models (LLMs) with external evidence by retrieving documents at inference time, but it also turns the retrieval corpus into a sensitive asset.
    Under a black-box setting, an adversary \emph{given a candidate document} can infer whether it has been ingested into the RAG knowledge base (\emph{i.e.}, document-level membership inference) solely from query--response interactions, thereby leaking corpus coverage and the existence of sensitive topics.
    Existing RAG-MIA methods either rely on ``soft'' signals such as semantic similarity, which often yield overlapping member/non-member score distributions and unstable thresholds, or employ explicit confirmation probes whose intent is conspicuous and thus prone to refusal and detection.
    We propose \alg, which converts verifiable hard evidence in the target document (\emph{e.g.}, fine-grained details, proper nouns/technical terms, definitional statements, metadata cues, and causal/constraint relations) into an exam with four objectively gradable question types (FB/SC/MC/T/F), and uses the aggregated exam score across multiple evidence-targeted questions as the membership signal.
    Experiments across multiple datasets and diverse RAG configurations demonstrate that \alg\ improves member/non-member separability in stringent settings while preserving natural, stealthy queries, and we further analyze the impact of question composition and exam length on attack effectiveness.
\end{abstract}


\input{B-1-Introduction}
\input{B-2-RelatedWork}

\input{B-4-Approach}

\input{B-5-Analysis}

\input{B-5-Experiments-A}

\input{B-7-Conclusion}

\section*{Acknowledgments}
This work was supported by the National Natural Science Foundation of China (No.62272198), 
in part by Guangdong Basic and Applied Basic Research Foundation under Grant (No. 2024A1515010121),
in part by the Special Funds for the Cultivation of Guangdong College Students’ Scientific and Technological Innovation (Climbing Program Special Funds) under Grant pdjh2025ak028,
in part by the Outstanding Innovative Talents Cultivation Funded Programs for Graduate Students of Jinan University under Grant 2025CXY336,
and in part by the Cybersecurity College Student Innovation Funding Program (Topsec Technologies Group Inc).

\bibliographystyle{IEEEtran}
\bibliography{Ref}

\begin{IEEEbiography}
    [{\includegraphics[width=1in,height=1.25in,clip,keepaspectratio]{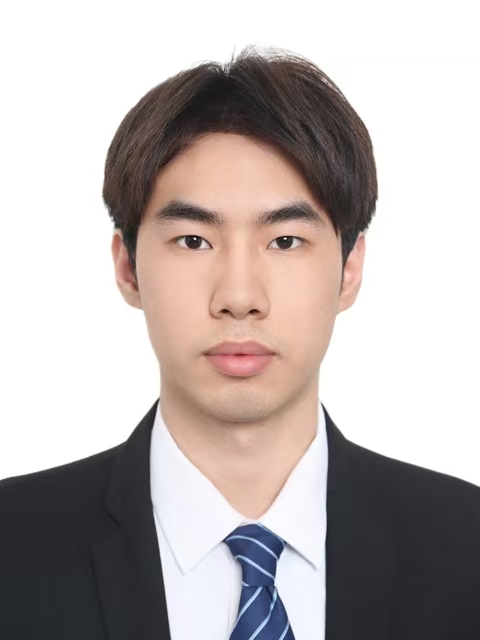}}]{Zelin Guan} received the B.A. degree in computer science and technology from Guangzhou University, Guangzhou, China in 2025. 
    He is currently pursuing the M.S. degree in Cyberspace Security at Jinan University, Guangzhou, China. 
    His research interests include Lager Language Model Security and Data Mining.
\end{IEEEbiography}

\begin{IEEEbiography}
    [{\includegraphics[width=1in,height=1.25in,clip,keepaspectratio]{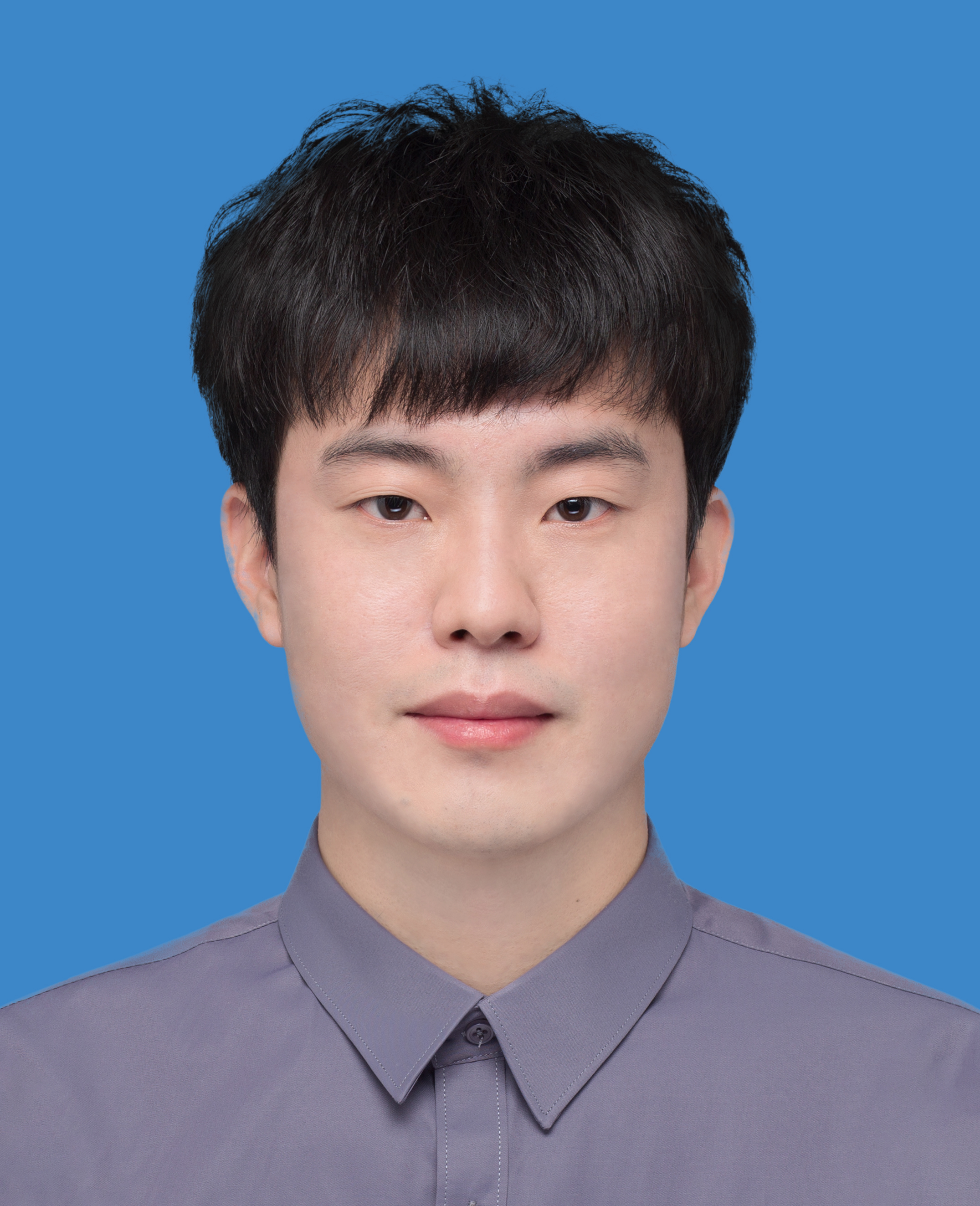}}]{Shengda Zhuo} (\emph{Student Member, IEEE}) received the M.S. degree in computer science and technology from Guangzhou University, Guangzhou, China in 2023. 
    He is currently pursuing the Ph.D. degree in Cyberspace Security at Jinan University, Guangzhou, China. 
    He is a joint Ph.D. student at The Hong Kong Polytechnic University (PloyU) from December 2024 to December 2025.
    His research interests include Machine Learning, Recommendation Systems, and Data Mining.
\end{IEEEbiography}

\begin{IEEEbiography}
    [{\includegraphics[width=1in,height=1.25in,clip,keepaspectratio]{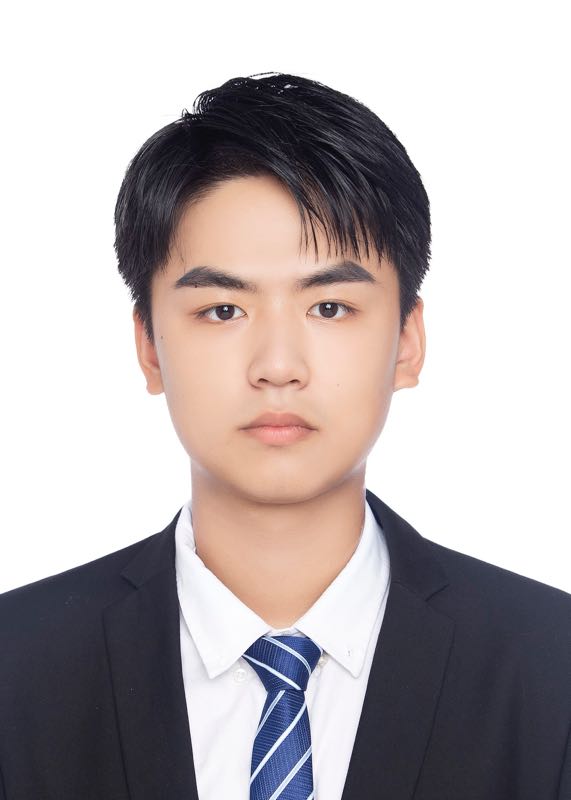}}]{Zeyan Li} (\emph{Student Member, IEEE}) is a dual-degree student in the Department of Information and Computational Science at Jinan University and the Department of Applied Mathematics and Informatics at the University of Birmingham. His research focuses on artificial intelligence and data mining. He will join the Department of Computer Science at Shanghai Jiao Tong University next year to pursue a Ph.D. in Electronic Information.
\end{IEEEbiography}

\begin{IEEEbiography}
    [{\includegraphics[width=1in,height=1.25in,clip,keepaspectratio]{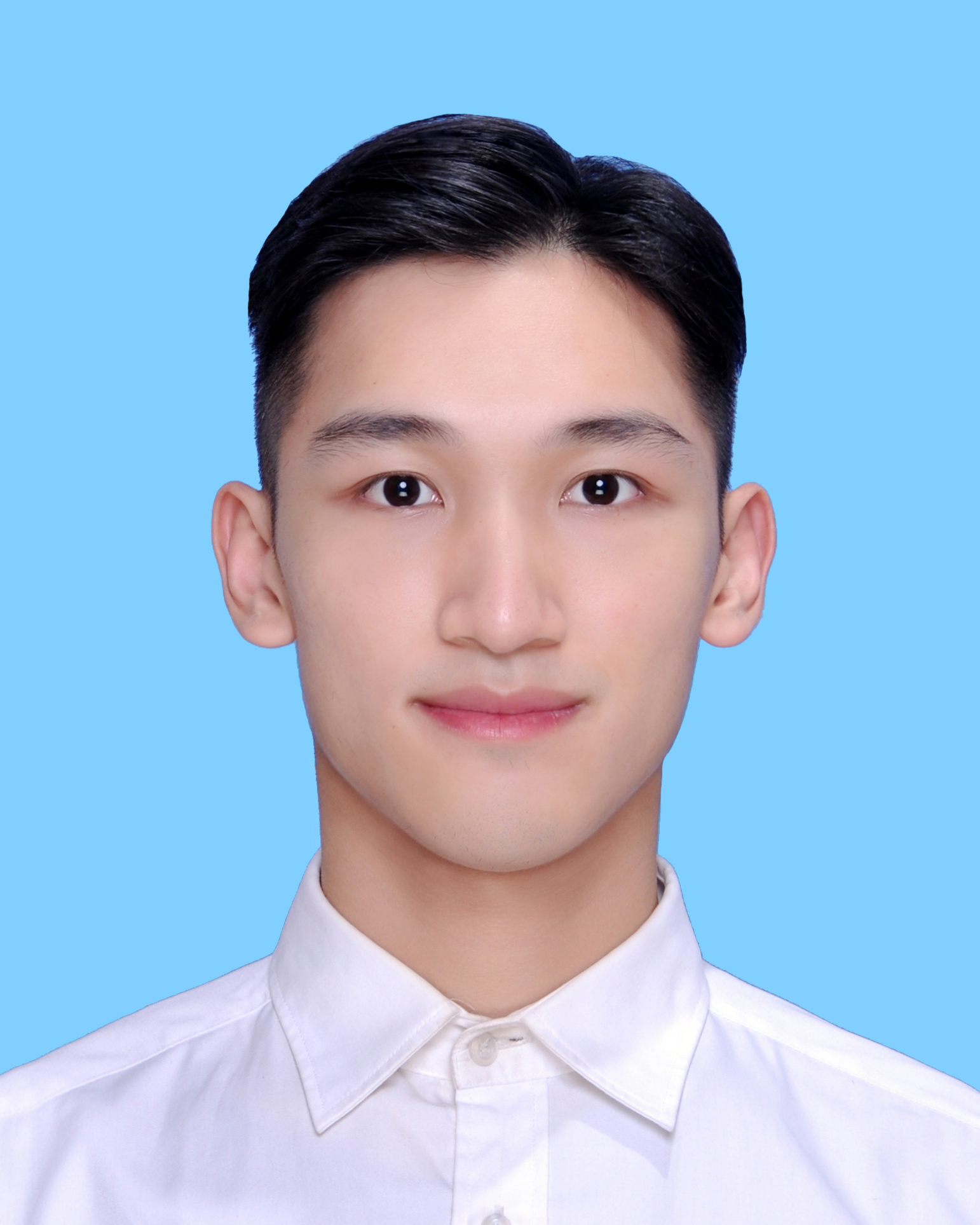}}]{Jinchun He} received the M.S. degree in computer technology from Guangzhou University, Guangzhou, China in 2023. He is currently pursuing the Ph.D degree in the Institute of Artificial Intelligence at Beihang University, Beijing, China. His research interests include blockchain, cybersecurity, and distributed computing.
\end{IEEEbiography}

\begin{IEEEbiography}
    [{\includegraphics[width=1in,height=1.25in,clip,keepaspectratio]{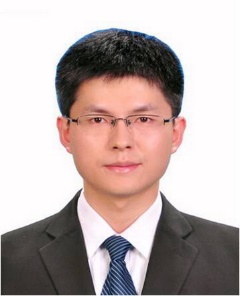}}]{Wangjie Qiu} is currently an associate professor with Beijing Advanced Innovation Center for Future Blockchain and Privacy Computing, Beihang University. He received his B.S. at 2008 and Ph.D degree at 2013, both in mathematics from Beihang University. His research interests include Information security, blockchain and privacy computing. And he is also the deputy secretary general of blockchain committee, China institute of communications. As the founding team, he successfully developed ChainMaker, an international advanced blockchain technology system. He has been authorized a number of invention patents in the fields of information security, blockchain and privacy computing and he won the first prize of China National Technology Invention in 2014.  
\end{IEEEbiography}

\begin{IEEEbiography}    
    [{\includegraphics[width=1in,height=1.25in,clip,keepaspectratio]{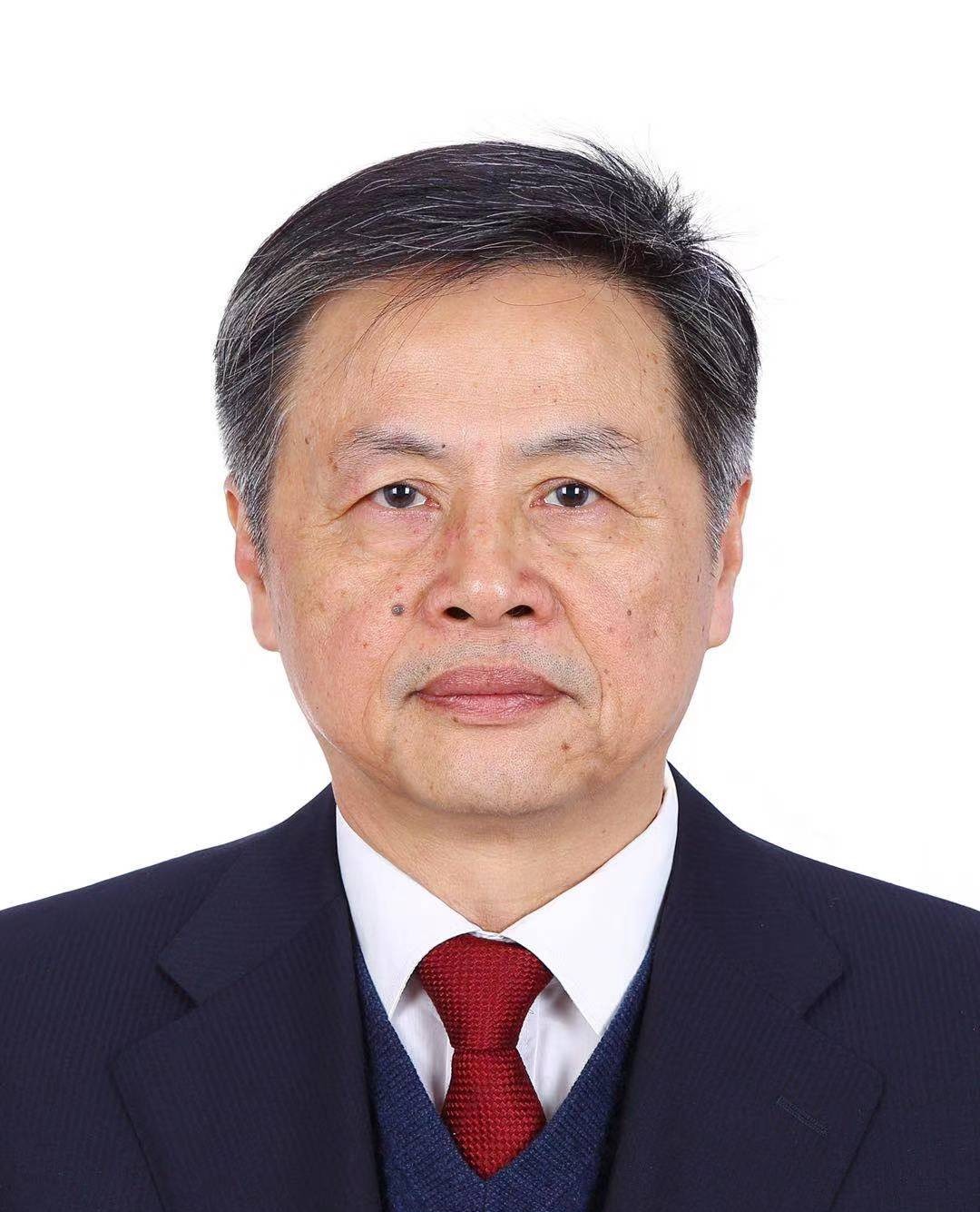}}]{Zhiming Zheng} received the Ph.D. degree in mathematics from the School of Mathematical Sciences, Peking University, Beijing, China, in 1987. He is currently a Professor with the Institute of Artificial Intelligence, Beihang University, Beijing, China. His research interests include refined intelligence, blockchain, and privacy computing. He is one of the initiators of Blockchain-ChainMaker. He is a member of \textbf{Chinese Academy of Sciences}.
\end{IEEEbiography}

\begin{IEEEbiography}    
  [{\includegraphics[width=1in,height=1.25in,
  clip,keepaspectratio]
  {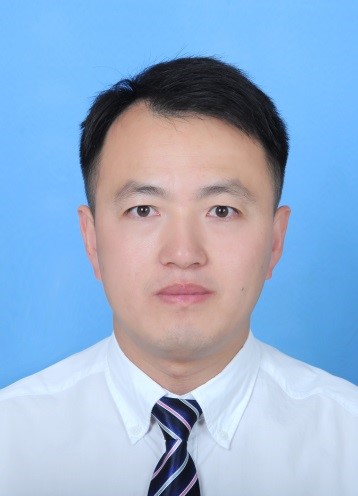}}]{Shuqiang Huang} received the Ph.D. degree in computer science in 2010 from South China University of Technology, Guangzhou, China. He is currently a full professor of college of cyber security of Jinan University. He is a distinguished member of China Computer Federation. His main research interests include edge computing, industrial IoT and artificial intelligence. He has published more than 50 academic papers in international journals such as IEEE TCYB, IEEE TPDS, IEEE TKDE, IEEE TII, TCS, and ACM TIST.
\end{IEEEbiography}

\end{document}

%% file: C-macro.tex
\usepackage{mathtools}
\usepackage{graphicx}
\usepackage[utf8]{inputenc} 
\usepackage[T1]{fontenc}    
\usepackage[noadjust]{cite}
\usepackage[bookmarks=true,breaklinks=true,colorlinks,linkcolor=RedViolet,citecolor=blue,urlcolor=BlueViolet]{hyperref}
\usepackage{url}            
\usepackage{booktabs,multirow}       
\usepackage{subfigure}
\usepackage{amsmath,amssymb,amsfonts,amsthm,stmaryrd}
\usepackage{mathtools}
\usepackage{nicefrac}       
\usepackage{microtype}      
\usepackage{authblk}
\usepackage[linesnumbered,ruled,vlined]{algorithm2e}
\usepackage{subcaption}
\usepackage{dblfloatfix}
\usepackage{dsfont}
\usepackage[misc]{ifsym} 
\usepackage[inline,shortlabels]{enumitem} 
\graphicspath{burl/Figure/}

\usepackage{siunitx}
\usepackage[dvipsnames]{xcolor}
\usepackage{pdfpages}

\renewcommand{\eqref}[1]{Eq.\,(\ref{#1})}

\newcommand{\yA}[1]{\hat{y}_{\text{obs}}}
\newcommand{\yR}[1]{\hat{y}_{\text{rec}}}

\newcommand{\romup}[1]{\uppercase\expandafter{\romannumeral #1\relax}}
\newcommand{\romlo}[1]{\lowercase\expandafter{\romannumeral #1\relax}}

\newcommand{\rom}[1]{\uppercase\expandafter{\romannumeral #1\relax}}
\newcommand{\romsm}[1]{\lowercase\expandafter{\romannumeral #1\relax}}

\renewenvironment{proof}{\noindent\textbf{\em proof:}}{\hfill$\square$}

\newcommand{\alg}{{BURL}} 

\usepackage[inline,shortlabels]{enumitem} 
\renewcommand{\eqref}[1]{Eq.\,(\ref{#1})}

%% file: B-1-Introduction.tex
\section{Introduction}
\label{sec:intro}
\IEEEPARstart{L}{arge} language models (LLMs)~\cite{anil2023palm,brown2020language,touvron2023llama} have demonstrated strong capabilities in open-domain question answering and knowledge-centric services. 
However, their generation is primarily grounded in static parametric knowledge, making it difficult to continuously incorporate up-to-date information, and they are prone to hallucinations~\cite{ji2023survey} when required evidence falls outside their internal parameters. 
Retrieval-Augmented Generation (RAG)~\cite{lewis2020retrieval,borgeaud2022improving} addresses these limitations by retrieving relevant documents from an external corpus prior to generation and injecting the resulting evidence into the context, thereby improving timeliness and traceability without frequent retraining. 
As a result, RAG has emerged as a mainstream solution for enterprises and organizations deploying knowledge-based question answering systems~\cite{diefenbach2018core}.


\begin{figure}[t!]
     \centering
     \includegraphics[width=1\linewidth]{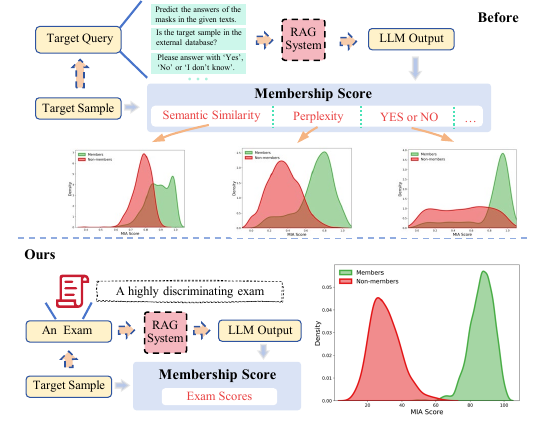}
     \caption{
     Traditional RAG-MIAs use soft signals (similarity/perplexity/yes-no), while our E-MIA uses aggregated exam scores for a more separable membership signal.
     }
     \label{fig:Compare}
\end{figure} 

RAG shifts key assets from model parameters to the external document corpus and its retrieval index, but this shift also introduces a new privacy leakage channel~\cite{krishnamurthy2011privacy, kim2023propile}: even without access to the knowledge base or retrieved contexts, an adversary who possesses a candidate document may still infer (solely from query-response interactions), whether the document has been ingested into the RAG corpus.
This document-level membership inference~\cite{meeus2024did, su2025falcon} risk is distinct from classical training-data membership inference: the target is not the training set, but the existence and coverage of the retrieval corpus. 
For knowledge bases containing internal policies, business records, or copyright-protected materials, even the binary fact of whether a document is present can raise compliance concerns and expose sensitive business intelligence.

Existing RAG membership inference attacks under black-box access still face two practical limitations. 
First, many approaches construct membership scores from semantic similarity~\cite{li2025generating,DBLP:conf/icissp/AndersonAG25} between the output and the target text (or a reference answer), or from generation-quality metrics. 
Yet, due to strong generalization and paraphrasing, member and non-member score distributions often overlap substantially, leading to unstable threshold-based decisions~\cite{hu2023defenses}.
Second, confirmation-style probes attempt to directly elicit whether a document exists in the corpus; while such probes can yield stronger signals, their intent is conspicuous and thus more likely to trigger refusals or detection mechanisms in real deployments~\cite{gospodinov2023doc2query}.
Some work further explores multi-query prompting or localized verification (\emph{e.g.}, masked completion) to validate fine-grained content, but these efforts largely remain within ``soft'' similarity signals or a single form of local check, and fall short of a unified framework that systematically leverages document-specific evidence while keeping prompts natural, objectively gradable, and statistically separable.

We advocate a more fundamental perspective: in black-box RAG, the feasibility of document-level membership inference is not primarily determined by increasingly sophisticated similarity metrics, but by whether the system can access document-specific, verifiable hard evidence.
In practice, a document often contains 
(1) \emph{precise details} (\emph{e.g.}, numbers, dates, conditions, and ordering), 
(2) \emph{proper names and domain-specific terminology}, 
(3) \emph{definitional statements} (\emph{e.g.}, crisp concept boundaries and exact definitions), 
(4) \emph{metadata cues} (\emph{e.g.}, titles, sections, identifiers, and versions), and 
(5) \emph{constraint relations} (\emph{e.g.}, premise--conclusion links and condition--action rules). 
If the document is ingested into the corpus, it becomes retrievable with non-trivial probability, making such evidence more likely to be reproduced consistently across multiple evidence-targeted questions; if the document is absent, the model must rely on generalization or guessing and is most prone to systematic errors on these verifiable points.
These properties make hard evidence a stronger basis for discrimination than global semantic similarity.
Fig.~\ref{fig:Compare} contrasts prior RAG-MIAs that score membership using soft signals (\emph{e.g.}, similarity/perplexity/yes-no) with our exam-based E-MIA that aggregates evidence-targeted exam scores, yielding a more separable decision signal in our evaluation~\cite{li2025generating,DBLP:conf/icissp/AndersonAG25,meeus2024did}.

Building on this insight, we propose \textbf{E-MIA} (Exam-style Membership Inference Attack), which reformulates membership inference as a systematic exam over document-specific evidence.
Rather than relying on ambiguous signals from a single query, \alg\ generates structured questions anchored to verifiable evidence points and uses the system's performance on the resulting question set as the membership score.
To balance natural prompting with objective grading, we adopt four basic question formats (\emph{i.e.}, single-choice (SC), multiple-choice (MC), true/false (T/F), and fill-in-the-blank (FB)), so that evidence verification can be posed in standard QA form while yielding stable, quantitative outcomes.
\alg\ derives its discriminative power from aggregating signals across multiple evidence points and multiple questions, which mitigates variance induced by retrieval mismatch and reduces both accidental hits and ``plausible'' generalized answers.
This aggregation widens the statistical gap between member and non-member score distributions, making thresholding more robust and membership inference more reliable.

\alg\ is designed for common constraints in real-world deployments: the adversary has only API-level black-box access, request filtering (\emph{e.g.}, malicious-input detectors), while being unable to observe retriever scores, retrieved contexts/candidate documents, or internal logs.
Accordingly, our design emphasizes three principles: \emph{robust black-box applicability} (scoring relies solely on model outputs), 
\emph{stealthiness} (prompts resemble standard reading-comprehension or knowledge-verification tasks and avoid explicit confirmation-style probing to reduce rejection risk), 
and \emph{efficiency} (stable decisions are obtained from aggregated exam scores under limited query budgets).
We summarize our \textbf{Contributions} as follows:
\begin{itemize}[leftmargin=*]
    \item We first introduce an evidence-centric view of RAG document membership inference by decomposing a target document into typed, verifiable hard-evidence units that admit objective grading.
    \item We propose \alg$\footnote{Code:\url{https://github.com/508954254/E-MIA}}$, a unified exam-style black-box framework that converts evidence into natural-looking questions across four formats and aggregates outcomes into a robust membership inference score.
    \item We conduct extensive and reproducible evaluations across multiple datasets and target models, systematically analyzing key factors (\emph{e.g.}, question-type mix, exam length, evidence categories, retrieval budget, and exam generator choice) that affect both attack effectiveness and stealthiness.
\end{itemize}

This paper is structured to provide a comprehensive understanding of our research. 
Section~\ref{sec:relatedwork} reviews the related work on retrieval-augmented generation and membership inference attacks. 
Section~\ref{sec:preliminiaries} introduces the preliminaries, including the attacker’s target, constraints, and the formulation of the exam-style membership inference task. 
Section~\ref{sec:document} presents the evidence-centric document decomposition, detailing how target documents are transformed into verifiable evidence units. 
Section~\ref{sec:framework} introduces the proposed E-MIA methodology, including exam-style question generation and the aggregation-based scoring mechanism. 
In Section~\ref{sec:experiments}, we conduct extensive experiments to evaluate the effectiveness and robustness of the proposed approach. 
Finally, Section~\ref{sec:conclusion} concludes the paper and discusses potential directions for future work.

%% file: B-2-RelatedWork.tex
\section{Related Work}
\label{sec:relatedwork}
Existing studies can be broadly categorized into two main lines: retrieval-augmented generation frameworks and membership inference attacks in RAG systems.

\subsection{Retrieval-Augmented Generation (RAG)}
Retrieval-Augmented Generation (RAG) mitigates knowledge staleness and hallucinations inherent to static parametric models by incorporating an external knowledge base at generation time, thereby providing large language models with retrievable evidence as contextual grounding~\cite{ji2023survey,brown2020language,touvron2023llama,gao2024retrievalaugmentedgenerationlargelanguage,arzanipour2025ragsecurityprivacyformalizing}. 
A typical RAG pipeline consists of a retriever and a generator: given a user query, the retriever returns the top-k relevant document passages from an external corpus, which are then concatenated with the query and fed into the generator to produce the final response. 
Because the model’s output is explicitly conditioned on the retrieved content, the external document repository and its associated index constitute core assets of a RAG system~\cite{anil2023palm,brown2020language,touvron2023llama}. 
This dependency also introduces a new privacy exposure surface: even if the system does not directly disclose verbatim text, an adversary may still infer (solely via black-box interactions and output analysis) whether a particular document is included in the corpus or is retrievable by the system.

\subsection{Membership Inference in RAG Systems (RAG-MIA)}
Membership inference attacks\cite{hu2022membership,choquette2021label,he2025towards} in the RAG setting (RAG-MIA) aim to determine whether a target document (or target sample) is present in the retrieval corpus/index of a RAG system, where the adversary typically has only black-box query access and can observe the system’s outputs~\cite{liu2025mask,naseh2025riddle,gao2025dcmi,li2025budgetleak,shokri2017membership,feng2025ragleak}. 
Existing approaches can be broadly categorized by the source of the membership signal into three classes: 
(i) explicit confirmation methods directly ask whether target content appears in the retrieved context (\emph{e.g.}, requiring a Yes/No response). 
While such signals can be strong, the attack intent is often conspicuous and therefore prone to detection or refusal~\cite{DBLP:conf/icissp/AndersonAG25}; 
(ii) similarity- or generation-quality scoring methods derive a membership score from semantic similarity between the output and a reference answer (or target text), perplexity, or related metrics. 
These methods are black-box friendly but can be confounded by model generalization in RAG settings~\cite{li2025generating,naseh2025riddle}; 
(iii) content probing methods amplify the discrepancy between members and non-members via masked completion of the target text, difficulty/perturbation calibration, or side-channel cues (\emph{e.g.}, generation budgets). 
Although these techniques may improve separability, they often rely on stronger assumptions (\emph{e.g.}, additional controllable interfaces, a reference system, or multi-round probing) and entail more complex execution pipelines~\cite{gao2025dcmi,liu2025mask,li2025budgetleak,wang2025rag}.

\begin{figure*}[!t]
  \centering
  \includegraphics[width=1\textwidth]{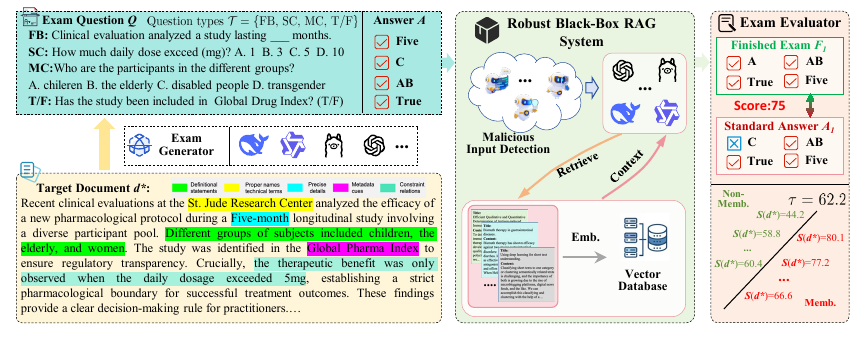}
  \caption{Overview of E-MIA, which infers document membership in a black-box RAG system by generating an exam from a target document, querying the RAG to complete it, and grading the answers to obtain a membership score
  }
  \label{fig:Overview}
\end{figure*}

Overall, practical RAG-MIA methods typically face a trade-off between robustness and feasibility. 
“Soft signals” based on semantic similarity or generation quality are susceptible to model generalization, leading to overlapping score distributions and unstable decision thresholds; conversely, enhancement strategies (\emph{e.g.}, masked completion and calibrated perturbations) can improve discriminability but frequently introduce extra assumptions (\emph{e.g.}, a reference model, controllable knobs, or multi-query overhead) and may be more likely to trigger defensive mechanisms~\cite{varshney2024art}. 
Departing from these paradigms, our work grounds membership inference in verifiable evidence intrinsic to the document by casting the task as an exam-style evaluation problem, using an aggregated exam score as the membership signal to achieve more stable statistical separation.

%% file: B-4-Approach.tex
\section{Preliminaries}
\label{sec:preliminiaries}
\par\noindent
\textbf{Attacker’s Target.}
We study document-level membership inference in RAG systems.
Given a target document $d^{*}$, the adversary aims to determine whether $d^{*}$ is contained in the RAG knowledge base $\mathcal{D}$, i.e., whether $d^{*}\in\mathcal{D}$ holds.

\par\noindent
\textbf{Attacker’s Constraints.}
We consider a realistic black-box setting: the adversary has no access to $\mathcal{D}$, retriever/LLM parameters, retrieved contexts $\mathcal{P}_k$, or retrieval scores, and may be filtered by a malicious-input detector.
The adversary can only issue queries and observe final textual outputs. We assume the adversary possesses the content of $d^{*}$ (as standard in membership inference), but cannot access the system’s KB or retrieval traces.
To improve feasibility and stealth, we craft prompts as natural QA or reading-comprehension questions rather than explicit confirmation-style probes.

\par\noindent
\textbf{Attacker’s Task.}
We instantiate the attack as an exam-style procedure.
The adversary generates an exam $X_{d^{*}}=\{(q_i,a_i)\}_{i=1}^{n}$, where $q_i$ is an exam question derived from evidence in $d^{*}$ and $a_i$ is its ground-truth answer.
Querying the RAG system yields responses $\{r_i\}_{i=1}^{n}$, and an \textit{Exam Evaluator} grades each response and aggregates results into an exam score: 
\begin{equation}
    S(d^{*})=\text{Agg}\big(\{\text{Score}(r_i,a_i)\}_{i=1}^{n}\big).
\end{equation}

Finally, the adversary outputs a membership decision $\hat{m}(d^{*})=\mathbb{I}[S(d^{*})\ge \tau]$.
In our design, $\text{Score}(\cdot)$ is deterministic and objectively checkable, and aggregation over multiple questions mitigates variance from retrieval mismatch.



\section{Evidence-Centric Document Decomposition}
\label{sec:document}
In document-level membership inference against black-box RAG systems, the adversary can only observe final responses and has no access to retrieved contexts, retrieval scores, or internal logs~\cite{meeus2024did,yang2025mrm}. 
As a result, signals based on global semantic similarity or coarse answer-quality assessments can be unstable: even when the target document is not retrieved, the model may still produce plausible answers via generalization and paraphrasing, leading to substantial overlap between member and non-member score distributions~\cite{gao2025dcmi}.
Motivated by this limitation, we shift the inference focus from overall similarity to whether the output consistently matches document-specific, verifiable hard evidence.
Intuitively, if a document is ingested into the knowledge base, it becomes retrievable with non-trivial probability, making the system more likely to reproduce its key facts and constraints across multiple queries; otherwise, deviations on those points occur more frequently.

To balance measurability with stealthiness, we require each evidence point to satisfy three criteria: it must admit objective grading, be discriminative of the target document, and be naturally expressible as conventional QA \cite{mansurova2024qa} or reading-comprehension style questions.
Under these criteria, the most discriminative evidence tends to fall into five categories: 
\begin{itemize}[leftmargin=*]
    \item \textbf{Precise details}: numbers, dates, thresholds, ordering, and conditional clauses, \emph{et al}.
    \item \textbf{Proper names and technical terms}: entities, acronyms, and domain-specific terminology, \emph{et al}.
    \item \textbf{Definitional statements}: crisp concept boundaries and exact definitions, \emph{et al}.
    \item \textbf{Metadata cues}: titles, section/clause identifiers, versions, and other structural markers, \emph{et al}.
    \item \textbf{Constraint relations}: premise--conclusion links, condition--action rules, and dependency chains, \emph{et al}.
\end{itemize}

These categories are not chosen for maximum coverage; rather, they are hard to guess reliably from “generalization-only” answers while remaining easy to grade deterministically.

Based on the above, we represent each document $d$ by an evidence set $\mathbf{E}(d^*)=\{e_j\}$, where each $e_j$ is typed and paired with a canonical answer and a grading rule.
We then map evidence units to objectively-gradable exam items (fill-in-the-blank, true/false, single-choice, and multiple-choice), enabling a systematic examination of document-specific evidence while keeping queries natural in appearance.

%% file: B-5-Analysis.tex
\section{E-MIA Methodology}
\label{sec:framework}
This section presents E-MIA, which reformulates document-level membership inference as an exam-style verification of evidence extracted from a target document. 
Specifically, as illustrated in Fig.~\ref{fig:Overview}, given an evidence set (\textbf{E}($d^*$)), we construct natural-language queries and generate a structured set of questions; the system’s performance on this question set is then evaluated via objective scoring to produce a stable membership inference score.
We provide detailed examples and prompt templates for our method in \textbf{Appendix~B-D}.

\subsection{Exam Query Construction}
Given $\mathbf{E}(d^{*})$, we construct a pool of candidate exam items $\mathcal{Q}$ by treating each evidence unit $e_j$ as the minimum generation granularity.
For every item, we retain essential metadata---including question type, prompt, (optional) answer choices, ground-truth answer, and scoring rule---so that it can be graded consistently from textual outputs only.
To enhance stealthiness, prompts are phrased as conventional reading-comprehension or fact-verification questions and avoid explicit confirmation wording.
We follow a minimal-exposure principle by revealing only short anchors (\emph{e.g.}, entities or a few keywords) rather than verbatim passages, and we support limited output variations via deterministic normalization and pattern extraction (\emph{e.g.}, option-letter extraction and numeric/date canonicalization).

\begin{figure}[!h]
  \centering
  \includegraphics[width=0.48\textwidth]{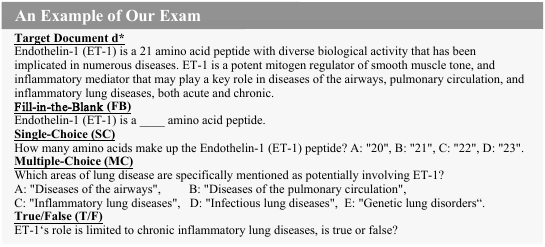}
  \caption{
  Example of an exam automatically generated from a target document $d^*$ (Endothelin-1), including four question types: Fill-in-the-Blank, Single-Choice, Multiple-Choice, and True/False
  }
  \label{fig:Example}
\end{figure}

Question instantiation follows a consistent evidence–to–question mapping. 
Precise details and proper nouns/technical terms are preferentially converted into single- or multiple-choices questions to enforce tight constraints. 
Definitional statements and metadata cues are better suited to fill-in-the-blank to preserve uniqueness. 
Local causal relations and constraint chains are validated via true/false or fill-in-the-blank questions that test necessary conditions and the directionality of relationships.
To mitigate accidental hits and generalization noise, the same evidence unit is often measured redundantly through multiple paraphrases and/or multiple formats, yielding a more stable statistical signal. 
We then apply quality filtering to remove questions with ambiguous answers, non-objective grading criteria, or insufficient stealthiness, and normalize answer representations (\emph{e.g.}, units, date formats, alias sets) to ensure consistent scoring and to facilitate subsequent exam assembly and score aggregation.
As illustrated in Fig.~\ref{fig:Example}, we instantiate a single target document into four objectively-gradable item types (FB, SC, MC, and T/F) while keeping prompts in a natural reading-comprehension style.

\subsection{Exam Generation}
Given the candidate exam pool ($\mathcal{Q}$), exam generation follows a three-stage pipeline: question-type specification, assembly, and validation.
We use membership separability, $\Delta = Acc_{\text{mem}} - Acc_{\text{non}}$, only for offline calibration on a held-out development set to guide design choices; the attack itself uses fixed specifications.
Based on this calibration (see \textbf{Sec.}~\ref{ablation and effectiveness studies} for mechanism analysis), we adopt the following default, reproducible rules for each question type:
\begin{itemize}[leftmargin=*]
    \item \textbf{Fill-in-the-Blank (FB):} We use one or two key blanks by default, which preserves minimal context while enforcing strong constraints.
    \item \textbf{Single-Choice (SC):} We use a small option set with a limited number of semantically similar distractors; specifically, $C \in \{4,5\}$ total options and $D \in \{1,2\}$ similar distractors.
    \item \textbf{Multiple-Choice (MC):} We use $n \in \{4,5\}$ options with the number of correct answers fixed to $m=2$.
    \item \textbf{True/False (T/F):} We generate items using fixed templates and consistent phrasing constraints to keep grading simple and stable.
\end{itemize}

After fixing type-specific specifications, we bucket questions by evidence category and question type, and sample $N$ questions (typically a multiple of 4 to control the type mix).
The sampling strategy prioritizes coverage of diverse evidence points, while allowing only a small subset of critical evidence units to be measured redundantly via paraphrases or cross-type variants to reduce variance from accidental hits and generalization noise.
We then perform consistency checks and cleaning, removing questions with multiple valid answers, those requiring overly long context for objective grading, and prompts whose wording may reveal explicit probing intent.
For choice-based items, distractors are constructed from nearby-but-incorrect candidates (\emph{e.g.}, close numerical values, similar entities, condition substitutions, or reversed causal directions) to avoid overly vague decision boundaries.
For FB questions, we enforce deterministic answer-format and normalization rules (\emph{e.g.}, units, date formats, alias sets) to ensure consistent grading based solely on the model output.
Through this assembly-and-validation procedure, we obtain a complete exam that is natural in appearance, objectively scorable, and reliably discriminative for subsequent inference scoring and threshold-based membership decisions.


\subsection{Inference Attack Scores}
To quantify membership, we normalize the exam outcome to a 0--100 \emph{membership score}.
Let $\mathcal{T}=\{\text{FB, SC, MC, T/F}\}$ denote the four question types, and let $X_i$ be the subset of exam items of type $i$.
We define the type-wise accuracy as:
\begin{equation}
    Acc_i = \frac{1}{|X_i|}\sum_{(q,a)\in X_i} \mathbb{I}\big[\text{Score}(r,a)=1\big],
\end{equation}
where $\text{Score}(\cdot)$ is deterministic and objectively checkable under our evaluator.
We aggregate type-wise accuracies using a weighted sum:
\begin{equation}
    S = 100 \sum_{i \in \mathcal{T}} w_i \cdot Acc_i,
\end{equation}
with $\sum_i w_i=1$. 
We typically set the total number of questions $N$ as a multiple of 4 to keep the type mix balanced.

We choose $w_i$ via offline calibration on a held-out development set by emphasizing question types whose member/non-member score distributions are more separable (measured by KL divergence), and keep $w_i$ fixed during attack.
We use the calibrated default weights reported in \textbf{Section~\ref{sec:Parameter Calibration}}.
Unless otherwise specified, we use the calibrated defaults $w_{\text{FB}}=0.312$, $w_{\text{SC}}=0.214$, $w_{\text{MC}}=0.300$, and $w_{\text{T/F}}=0.174$.

Finally, we output a binary membership decision with a threshold:
\begin{equation}
    \hat{m}(d^{*})=\mathbb{I}[S(d^{*})\ge \tau],
\end{equation}
where we select $\tau$ on the same calibration set to maximize balanced accuracy (\emph{cf}.\ Fig.~\ref{fig:44}), and use the default $\tau=62.2$ in all experiments.

%% file: B-5-Experiments-A.tex
\section{Experiments}
\label{sec:experiments}
In this section, we conduct extensive experiments on three real-world datasets to validate and evaluate the performance of the proposed model. Specifically, we aim to investigate the following five research questions (RQs):

\begin{itemize}[leftmargin=*]
    \item \textbf{RQ1:} How effective is \alg\ compared to existing MIA methods?
    \item \textbf{RQ2:} Why does the proposed exam-based design improve membership inference?
    \item \textbf{RQ3:} Is \alg\ robust against practical defense mechanisms and system variations?
    \item \textbf{RQ4:} How efficient is \alg\ in terms of parameter sensitivity and query budget?
    \item \textbf{RQ5:} Does \alg\ generalize across different models, retrievers, and languages?
\end{itemize}

\subsection{Experiment Details}
\par\noindent
\textbf{Datasets.} 
\label{sec:datasets}
To evaluate the robustness of our attack across corpora of different scales, we select three datasets from the BEIR benchmark~\cite{thakur2021beir}: NFCorpus$\footnote{NFCorpus: https://huggingface.co/datasets/BeIR/nfcorpus/}$, TREC-COVID$\footnote{TREC-COVID: https://huggingface.co/datasets/BeIR/trec-covid/}$, and SCIDOCS$\footnote{SCIDOCS: https://huggingface.co/datasets/BeIR/scidocs/}$.
They cover scientific and biomedical domains and contain approximately 3.5K, 23K, and 116K documents, respectively.
Table~\ref{tab:datasets} summarizes the datasets we used.
We apply a unified preprocessing pipeline to each corpus: we remove documents shorter than 50 tokens to ensure sufficient semantic content and eliminate duplicates to avoid evaluation bias.
After preprocessing, we randomly split each corpus by designating 60\% of the documents as members (\emph{i.e.}, ingested into the RAG knowledge base) and the remaining 40\% as non-members.
For each dataset, we then construct a balanced evaluation set of 2,000 documents (\emph{i.e.}, 1,000 randomly sampled members and 1,000 randomly sampled non-members) to assess attack performance.
A target document is labeled as \emph{member} if it is ingested into the RAG knowledge base. 
\emph{Non-member} targets are held out from the knowledge base by construction, ensuring that any correct evidence must come from memorization or spurious correlations rather than direct retrieval.

\begin{table}[!t]
    \centering
    \caption{Characteristics of the studied datasets.}
    \setlength{\tabcolsep}{6pt}
    \renewcommand{\arraystretch}{0.85}
    \scalebox{0.9}{
    \label{tab:datasets}
    \begin{tabular}{m{1.2cm}|m{1cm}|m{0.9cm}|m{0.9cm}|m{0.4\linewidth}}
        \toprule
        \midrule
        \textbf{Datasets} & \textbf{Domain} & \textbf{Number} & \textbf{Avg Length} & \textbf{Example} \\
        \midrule
        NFCorpus & Bio-Medical & 3.5K & 1496.92 & Recent studies have suggested that statins, an established drug group in the prevention of cardiovascular mortality...  \\
        \midrule
        TREC-COVID & Bio-Medical & 23K & 1356.24 & This retrospective chart review describes the epidemiology and clinical features of 40 patients with culture-proven Mycoplasma pneumoniae... \\
        \midrule
        SCIDOCS & Science & 116K & 1147.23 & An evolutionary recurrent network which automates the design of recurrent neural/fuzzy networks using a new evolutionary learning algorithm... \\
        \bottomrule
    \end{tabular}
    }
\end{table}

\par\noindent
{\bf Exam Generator.} 
\label{Generator}
We use the DeepSeek-R1~\cite{guo2025deepseek} API as the generation engine and design question-type-specific prompt templates to guide the model in producing exam questions tailored to different attack vectors (see \textbf{Appendix~C}). 
This design enables systematic probing of the target document's membership signal.

\par\noindent
\textbf{Exam Evaluator.}
\label{Exam Evaluator}
We evaluate objective questions using an automated scoring framework centered on regular-expression (regex) pattern matching against gold-standard templates. 
Raw model outputs undergo a normalization stage, removing conversational noise, redundant whitespace, and formatting artifacts, before the evaluator applies normalized exact match for FB and T/F tasks. 
For MC and SC, a regex-based extraction protocol isolates specific option identifiers (\emph{e.g.}, A-D) and utilizes set-based comparison to verify accuracy, ensuring a scoring process is strictly objective. (see \textbf{Appendix~D}) 

\begin{table*}[!t]
    \centering
    \caption{Comparison of membership inference attacks across three RAG corpora and four LLM backbones \hl{(Deepseek-R1-14B, Llama3.1-8B, Qwen2.5-7B, and Gemma2-2B)}, reporting Acc, AUC-ROC, AUC-PR, and TPR@FPR (0.05/0.01/0.005).
    }
    \label{tab:overall}
    \setlength{\tabcolsep}{8pt}
    \renewcommand{\arraystretch}{1}
    \scalebox{0.9}{
    \begin{tabular}{ccc|cccccc}
        \toprule
        \midrule
        \multirow{2}{*}[0pt]{\textbf{Dataset}} & \multirow{2}{*}[0pt]{\textbf{Models}} & \multirow{2}{*}[0pt]{\textbf{Attack Method}} & \multirow{2}{*}[0pt]{\textbf{Accuracy}} & \multirow{2}{*}[0pt]{\textbf{AUC-ROC}} & \multirow{2}{*}[0pt]{\textbf{AUC-PR}} & \multicolumn{3}{c}{\textbf{TPR@FPR}} \\
        & & & & & & \textbf{FPR=0.05} & \textbf{FPR=0.01} & \textbf{FPR=0.005} \\ 
        \midrule
        \multirow{21}{*}[-8pt]{\textbf{NFCorpus}} & \multirow{4}{*}[0pt]{\textbf{Deepseek-r1-14B}} & S$^2$MIA & 0.714 & 0.796 & 0.833 & 0.487 & 0.357 & 0.314 \\
        & & MBA & 0.726 & 0.805 & 0.841 & 0.468 & 0.354 & 0.309 \\
        & & IA & 0.925 & 0.914 & 0.915 & 0.816 & 0.523 & 0.254 \\
        & & E-MIA(Ours) & \raisebox{-1pt}{\textbf{\small 0.977}} & \raisebox{-1pt}{\textbf{\small 0.995}} & \raisebox{-1pt}{\textbf{\small 0.994}} & \raisebox{-1pt}{\textbf{\small 0.993}} & \raisebox{-1pt}{\textbf{\small 0.911}} & \raisebox{-1pt}{\textbf{\small 0.817}}\\
        & & \cellcolor{blue!12}\raisebox{-1pt}{\emph{Improve}} & \cellcolor{blue!12}\raisebox{-1.25pt}{5.6\%$\uparrow$} & \cellcolor{blue!12}\raisebox{-1.25pt}{8.1\%$\uparrow$} & \cellcolor{blue!12}\raisebox{-1.25pt}{8.6\%$\uparrow$} & \cellcolor{blue!12}\raisebox{-1.25pt}{21.6\%$\uparrow$} & \cellcolor{blue!12}\raisebox{-1.25pt}{74.2\%$\uparrow$} & \cellcolor{blue!12}\raisebox{-1.25pt}{160.2\%$\uparrow$} \\
        
        \cmidrule{3-9}
        & \multirow{4}{*}[0pt]{\textbf{Llama3.1-8B}} & S$^2$MIA & 0.676 & 0.802 & 0.838 & 0.485 & 0.365 & 0.332 \\
        & & MBA & 0.734 & 0.842 & 0.864 & 0.558 & 0.432 & 0.338 \\
        & & IA & 0.932 & 0.929 & 0.913 & 0.967 & 0.827 & 0.505 \\
        & & E-MIA(Ours) & \raisebox{-1pt}{\textbf{\small 0.995}} & \raisebox{-1pt}{\textbf{\small 0.996}} & \raisebox{-1pt}{\textbf{\small 0.989}} & \raisebox{-1pt}{\textbf{\small 1.000}} & \raisebox{-1pt}{\textbf{\small 0.998}} & \raisebox{-1pt}{\textbf{\small 0.976}} \\
        & & \cellcolor{blue!12}\raisebox{-1pt}{\emph{Improve}} & \cellcolor{blue!12}\raisebox{-1.25pt}{6.8\%$\uparrow$} & \cellcolor{blue!12}\raisebox{-1.25pt}{5.8\%$\uparrow$} & \cellcolor{blue!12}\raisebox{-1.25pt}{8.3\%$\uparrow$} & \cellcolor{blue!12}\raisebox{-1.25pt}{3.4\%$\uparrow$} & \cellcolor{blue!12}\raisebox{-1.25pt}{20.7\%$\uparrow$} & \cellcolor{blue!12}\raisebox{-1.25pt}{93.3\%$\uparrow$} \\
        
        \cmidrule{3-9}
        & \multirow{4}{*}[0pt]{\textbf{Qwen2.5-7B}} & S$^2$MIA & 0.741 & 0.834 & 0.868 & 0.555 & 0.447 & 0.387 \\
        & & MBA & 0.757 & 0.835 & 0.878 & 0.558 & 0.455 & 0.322 \\
        & & IA & 0.932 & 0.928 & 0.932 & 0.917 & 0.752 & 0.474 \\
        & & E-MIA(Ours) & \raisebox{-1pt}{\textbf{\small 0.990}} & \raisebox{-1pt}{\textbf{\small 0.982}} & \raisebox{-1pt}{\textbf{\small 0.986}} & \raisebox{-1pt}{\textbf{\small 0.985}} & \raisebox{-1pt}{\textbf{\small 0.984}} & \raisebox{-1pt}{\textbf{\small 0.984}} \\
        & & \cellcolor{blue!12}\raisebox{-1pt}{\emph{Improve}} & \cellcolor{blue!12}\raisebox{-1.25pt}{6.2\%$\uparrow$} & \cellcolor{blue!12}\raisebox{-1.25pt}{5.8\%$\uparrow$} & \cellcolor{blue!12}\raisebox{-1.25pt}{5.8\%$\uparrow$} & \cellcolor{blue!12}\raisebox{-1.25pt}{7.4\%$\uparrow$} & \cellcolor{blue!12}\raisebox{-1.25pt}{30.9\%$\uparrow$} & \cellcolor{blue!12}\raisebox{-1.25pt}{107.6\%$\uparrow$} \\
        \cmidrule{3-9}
        & \multirow{4}{*}[0pt]{\textbf{Gemma2-2B}} & S$^2$MIA & 0.660 & 0.773 & 0.814 & 0.454 & 0.330 & 0.291 \\
        & & MBA & 0.627 & 0.723 & 0.711 & 0.357 & 0.254 & 0.229 \\
        & & IA & 0.929 & 0.926 & 0.921 & 0.947 & 0.612 & 0.401 \\
        & & E-MIA(Ours) & \raisebox{-1pt}{\textbf{\small 0.990}} & \raisebox{-1pt}{\textbf{\small 0.997}} & \raisebox{-1pt}{\textbf{\small 0.994}} & \raisebox{-1pt}{\textbf{\small 1.000}} & \raisebox{-1pt}{\textbf{\small 0.988}} & \raisebox{-1pt}{\textbf{\small 0.842}} \\
        & & \cellcolor{blue!12}\raisebox{-1pt}{\emph{Improve}} & \cellcolor{blue!12}\raisebox{-1.25pt}{6.6\%$\uparrow$} & \cellcolor{blue!12}\raisebox{-1.25pt}{7.7\%$\uparrow$} & \cellcolor{blue!12}\raisebox{-1.25pt}{7.9\%$\uparrow$} & \cellcolor{blue!12}\raisebox{-1.25pt}{5.6\%$\uparrow$} & \cellcolor{blue!12}\raisebox{-1.25pt}{61.4\%$\uparrow$} & \cellcolor{blue!12}\raisebox{-1.25pt}{110.1\%$\uparrow$} \\
        \cmidrule{3-9}
        & & \cellcolor{red!12}\raisebox{-1pt}{\emph{AVG Improve$\uparrow$}} & \cellcolor{red!12}\raisebox{-1.25pt}{6.3\%$\uparrow$} & \cellcolor{red!12}\raisebox{-1.25pt}{6.9\%$\uparrow$} & \cellcolor{red!12}\raisebox{-1.25pt}{7.7\%$\uparrow$} & \cellcolor{red!12}\raisebox{-1.25pt}{9.5\%$\uparrow$} & \cellcolor{red!12}\raisebox{-1.25pt}{46.8\%$\uparrow$} & \cellcolor{red!12}\raisebox{-1.25pt}{117.8\%$\uparrow$} \\
        \midrule
        \multirow{21}{*}[-8pt]{\textbf{TREC-COVID}} & \multirow{4}{*}[0pt]{\textbf{Deepseek-r1-14B}} & S$^2$MIA & 0.749 & 0.829 & 0.865 & 0.546 & 0.528 & 0.377 \\
        & & MBA & 0.785 & 0.856 & 0.869 & 0.568 & 0.514 & 0.362 \\
        & & IA & 0.876 & 0.915 & 0.895 & 0.877 & 0.504 & 0.208 \\
        & & E-MIA(Ours) & \raisebox{-1pt}{\textbf{\small 0.989}} & \raisebox{-1pt}{\textbf{\small 0.998}} & \raisebox{-1pt}{\textbf{\small 0.997}} & \raisebox{-1pt}{\textbf{\small 0.999}} & \raisebox{-1pt}{\textbf{\small 0.997}} & \raisebox{-1pt}{\textbf{\small 0.837}} \\
        & & \cellcolor{blue!12}\raisebox{-1pt}{\emph{Improve}} & \cellcolor{blue!12}\raisebox{-1.25pt}{12.9\%$\uparrow$} & \cellcolor{blue!12}\raisebox{-1.25pt}{9.1\%$\uparrow$} & \cellcolor{blue!12}\raisebox{-1.25pt}{11.4\%$\uparrow$} & \cellcolor{blue!12}\raisebox{-1.25pt}{13.9\%$\uparrow$} & \cellcolor{blue!12}\raisebox{-1.25pt}{97.8\%$\uparrow$} & \cellcolor{blue!12}\raisebox{-1.25pt}{122.0\%$\uparrow$} \\
        \cmidrule{3-9}
        & \multirow{4}{*}[0pt]{\textbf{Llama3.1-8B}} & S$^2$MIA & 0.732 & 0.833 & 0.869 & 0.557 & 0.461 & 0.393 \\
        & & MBA & 0.745 & 0.842 & 0.852 & 0.574 & 0.472 & 0.373 \\
        & & IA & 0.915 & 0.952 & 0.935 & 0.897 & 0.416 & 0.393 \\
        & & E-MIA(Ours) & \raisebox{-1pt}{\textbf{\small 0.991}} & \raisebox{-1pt}{\textbf{\small 0.999}} & \raisebox{-1pt}{\textbf{\small 0.999}} & \raisebox{-1pt}{\textbf{\small 0.999}} & \raisebox{-1pt}{\textbf{\small 0.992}} & \raisebox{-1pt}{\textbf{\small 0.982}} \\
        & & \cellcolor{blue!12}\raisebox{-1pt}{\emph{Improve}} & \cellcolor{blue!12}\raisebox{-1.25pt}{12.9\%$\uparrow$} & \cellcolor{blue!12}\raisebox{-1.25pt}{4.9\%$\uparrow$} & \cellcolor{blue!12}\raisebox{-1.25pt}{6.8\%$\uparrow$} & \cellcolor{blue!12}\raisebox{-1.25pt}{11.4\%$\uparrow$} & \cellcolor{blue!12}\raisebox{-1.25pt}{138.5\%$\uparrow$} & \cellcolor{blue!12}\raisebox{-1.25pt}{149.8\%$\uparrow$} \\
        \cmidrule{3-9}
        & \multirow{4}{*}[0pt]{\textbf{Qwen2.5-7B}} & S$^2$MIA & 0.767 & 0.845 & 0.885 & 0.611 & 0.518 & 0.462 \\
        & & MBA & 0.761 & 0.838 & 0.883 & 0.456 & 0.434 & 0.353\\
        & & IA & 0.942 & 0.938 & 0.956 & 0.899 & 0.607 & 0.406 \\
        & & E-MIA(Ours) & \raisebox{-1pt}{\textbf{\small 0.996}} & \raisebox{-1pt}{\textbf{\small 1.000}} & \raisebox{-1pt}{\textbf{\small 1.000}} & \raisebox{-1pt}{\textbf{\small 1.000}} & \raisebox{-1pt}{\textbf{\small 0.997}} & \raisebox{-1pt}{\textbf{\small 0.996}} \\
        & & \cellcolor{blue!12}\raisebox{-1pt}{\emph{Improve}} & \cellcolor{blue!12}\raisebox{-1.25pt}{8.4\%$\uparrow$} & \cellcolor{blue!12}\raisebox{-1.25pt}{6.6\%$\uparrow$} & \cellcolor{blue!12}\raisebox{-1.25pt}{4.6\%$\uparrow$} & \cellcolor{blue!12}\raisebox{-1.25pt}{11.2\%$\uparrow$} & \cellcolor{blue!12}\raisebox{-1.25pt}{60.9\%$\uparrow$} & \cellcolor{blue!12}\raisebox{-1.25pt}{115.6\%$\uparrow$} \\
        \cmidrule{3-9}
        & \multirow{4}{*}[0pt]{\textbf{Gemma2-2B}} & S$^2$MIA & 0.703 & 0.792 & 0.833 & 0.486 & 0.352 & 0.303 \\
        & & MBA & 0.687 & 0.756 & 0.784 & 0.417 & 0.302 & 0.218 \\
        & & IA & 0.932 & 0.922 & 0.955 & 0.827 & 0.533 & 0.356 \\
        & & E-MIA(Ours) & \raisebox{-1pt}{\textbf{\small 0.992}} & \raisebox{-1pt}{\textbf{\small 0.999}} & \raisebox{-1pt}{\textbf{\small 0.999}} & \raisebox{-1pt}{\textbf{\small 0.998}} & \raisebox{-1pt}{\textbf{\small 0.991}} & \raisebox{-1pt}{\textbf{\small 0.988}} \\
        & & \cellcolor{blue!12}\raisebox{-1pt}{\emph{Improve}} & \cellcolor{blue!12}\raisebox{-1.25pt}{6.9\%$\uparrow$} & \cellcolor{blue!12}\raisebox{-1.25pt}{8.4\%$\uparrow$} & \cellcolor{blue!12}\raisebox{-1.25pt}{4.6\%$\uparrow$} & \cellcolor{blue!12}\raisebox{-1.25pt}{20.7\%$\uparrow$} & \cellcolor{blue!12}\raisebox{-1.25pt}{85.9\%$\uparrow$} & \cellcolor{blue!12}\raisebox{-1.25pt}{177.5\%$\uparrow$} \\
        \cmidrule{3-9}
        & & \cellcolor{red!12}\raisebox{-1pt}{\emph{AVG Improve$\uparrow$}} & \cellcolor{red!12}\raisebox{-1.25pt}{10.3\%$\uparrow$} & \cellcolor{red!12}\raisebox{-1.25pt}{7.3\%$\uparrow$} & \cellcolor{red!12}\raisebox{-1.25pt}{6.9\%$\uparrow$} & \cellcolor{red!12}\raisebox{-1.25pt}{14.4\%$\uparrow$} & \cellcolor{red!12}\raisebox{-1.25pt}{95.8\%$\uparrow$} & \cellcolor{red!12}\raisebox{-1.25pt}{141.2\%$\uparrow$} \\
        \midrule
        \multirow{21}{*}[-8pt]{\textbf{SCIDOCS}} & \multirow{4}{*}[0pt]{\textbf{Deepseek-r1-14B}} & S$^2$MIA & 0.759 & 0.834 & 0.864 & 0.510 & 0.366 & 0.338 \\
        & & MBA & 0.762 & 0.808 & 0.824 & 0.506 & 0.426 & 0.361 \\
        & & IA & 0.935 & 0.950 & 0.958 & 0.923 & 0.684 & 0.397 \\
        & & E-MIA(Ours) & \raisebox{-1pt}{\textbf{\small 0.985}} & \raisebox{-1pt}{\textbf{\small 0.998}} & \raisebox{-1pt}{\textbf{\small 0.999}} & \raisebox{-1pt}{\textbf{\small 0.997}} & \raisebox{-1pt}{\textbf{\small 0.976}} & \raisebox{-1pt}{\textbf{\small 0.954}} \\
        & & \cellcolor{blue!12}\raisebox{-1pt}{\emph{Improve}} & \cellcolor{blue!12}\raisebox{-1.25pt}{5.3\%$\uparrow$} & \cellcolor{blue!12}\raisebox{-1.25pt}{5.1\%$\uparrow$} & \cellcolor{blue!12}\raisebox{-1.25pt}{4.3\%$\uparrow$} & \cellcolor{blue!12}\raisebox{-1.25pt}{8.0\%$\uparrow$} & \cellcolor{blue!12}\raisebox{-1.25pt}{42.7\%$\uparrow$} & \cellcolor{blue!12}\raisebox{-1.25pt}{140.3\%$\uparrow$} \\
        \cmidrule{3-9}
        & \multirow{4}{*}[0pt]{\textbf{Llama3.1-8B}} & S$^2$MIA & 0.746 & 0.832 & 0.864 & 0.523 & 0.395 & 0.345 \\
        & & MBA & 0.758 & 0.834 & 0.876 & 0.592 & 0.403 & 0.384 \\
        & & IA & 0.950 & 0.944 & 0.956 & 0.927 & 0.405 & 0.392 \\
        & & E-MIA(Ours) & \raisebox{-1pt}{\textbf{\small 0.997}} & \raisebox{-1pt}{\textbf{\small 0.999}} & \raisebox{-1pt}{\textbf{\small 0.999}} & \raisebox{-1pt}{\textbf{\small 0.999}} & \raisebox{-1pt}{\textbf{\small 0.998}} & \raisebox{-1pt}{\textbf{\small 0.997}} \\
        & & \cellcolor{blue!12}\raisebox{-1pt}{\emph{Improve}} & \cellcolor{blue!12}\raisebox{-1.25pt}{4.9\%$\uparrow$} & \cellcolor{blue!12}\raisebox{-1.25pt}{5.8\%$\uparrow$} & \cellcolor{blue!12}\raisebox{-1.25pt}{4.5\%$\uparrow$} & \cellcolor{blue!12}\raisebox{-1.25pt}{7.8\%$\uparrow$} & \cellcolor{blue!12}\raisebox{-1.25pt}{146.4\%$\uparrow$} & \cellcolor{blue!12}\raisebox{-1.25pt}{154.3\%$\uparrow$} \\
        \cmidrule{3-9}
        & \multirow{4}{*}[0pt]{\textbf{Qwen2.5-7B}} & S$^2$MIA & 0.782 & 0.859 & 0.890 & 0.596 & 0.487 & 0.434 \\
        & & MBA & 0.779 & 0.862 & 0.896 & 0.595 & 0.471 & 0.432 \\
        & & IA & 0.896 & 0.923 & 0.937 & 0.788 & 0.393 & 0.393 \\
        & & E-MIA(Ours) & \raisebox{-1pt}{\textbf{\small 0.999}} & \raisebox{-1pt}{\textbf{\small 0.999}} & \raisebox{-1pt}{\textbf{\small 1.000}} & \raisebox{-1pt}{\textbf{\small 0.999}} & \raisebox{-1pt}{\textbf{\small 0.998}} & \raisebox{-1pt}{\textbf{\small 0.998}} \\
        & & \cellcolor{blue!12}\raisebox{-1pt}{\emph{Improve}} & \cellcolor{blue!12}\raisebox{-1.25pt}{11.5\%$\uparrow$} & \cellcolor{blue!12}\raisebox{-1.25pt}{8.2\%$\uparrow$} & \cellcolor{blue!12}\raisebox{-1.25pt}{6.7\%$\uparrow$} & \cellcolor{blue!12}\raisebox{-1.25pt}{26.8\%$\uparrow$} & \cellcolor{blue!12}\raisebox{-1.25pt}{104.9\%$\uparrow$} & \cellcolor{blue!12}\raisebox{-1.25pt}{130.0\%$\uparrow$} \\
        \cmidrule{3-9}
        & \multirow{4}{*}[0pt]{\textbf{Gemma2-2B}} & S$^2$MIA & 0.722 & 0.803 & 0.839 & 0.474 & 0.352 & 0.334 \\
        & & MBA & 0.640 & 0.678 & 0.709 & 0.352 & 0.277 & 0.248 \\
        & & IA & 0.931 & 0.912 & 0.915 & 0.920 & 0.695 & 0.405 \\
        & & E-MIA(Ours) & \raisebox{-1pt}{\textbf{\small 0.996}} & \raisebox{-1pt}{\textbf{\small 0.999}} & \raisebox{-1pt}{\textbf{\small 0.999}} & \raisebox{-1pt}{\textbf{\small 0.999}} & \raisebox{-1pt}{\textbf{\small 0.997}} & \raisebox{-1pt}{\textbf{\small 0.996}} \\
        & & \cellcolor{blue!12}\raisebox{-1pt}{\emph{Improve}} & \cellcolor{blue!12}\raisebox{-1.25pt}{7.0\%$\uparrow$} & \cellcolor{blue!12}\raisebox{-1.25pt}{9.5\%$\uparrow$} & \cellcolor{blue!12}\raisebox{-1.25pt}{9.1\%$\uparrow$} & \cellcolor{blue!12}\raisebox{-1.25pt}{8.6\%$\uparrow$} & \cellcolor{blue!12}\raisebox{-1.25pt}{43.4\%$\uparrow$} & \cellcolor{blue!12}\raisebox{-1.25pt}{145.9\%$\uparrow$} \\
        \cmidrule{3-9}
        & & \cellcolor{red!12}\raisebox{-1pt}{\emph{AVG Improve$\uparrow$}} & \cellcolor{red!12}\raisebox{-1.25pt}{7.2\%$\uparrow$} & \cellcolor{red!12}\raisebox{-1.25pt}{7.2\%$\uparrow$} & \cellcolor{red!12}\raisebox{-1.25pt}{6.2\%$\uparrow$} & \cellcolor{red!12}\raisebox{-1.25pt}{9.0\%$\uparrow$} & \cellcolor{red!12}\raisebox{-1.25pt}{84.4\%$\uparrow$} & \cellcolor{red!12}\raisebox{-1.25pt}{142.6\%$\uparrow$} \\
        \midrule
        \bottomrule
    \end{tabular}
    }
\end{table*}

\par\noindent
{\bf RAG Settings.} 
\label{RAG Settings.}
We implement a standard RAG pipeline to systematically evaluate the effectiveness of \alg\ under diverse system configurations.
Our setup covers multiple combinations of retrievers and generators.
For generators, we select LLMs with different parameter scales to assess generalizability, including Gemma-2-2B~\cite{team2024gemma}, Qwen-2.5-7B~\cite{yang2024qwen2}, Llama-3.1-8B~\cite{dubey2024llama}, and DeepSeek-R1-14B~\cite{guo2025deepseek}.
For retrievers, we use BGE~\cite{zhang2023retrieve}, MiniLM-L6-v2~\cite{wang2020minilm}, and text-embedding-3-small~\cite{zhang2025qwen3} to span different embedding spaces~(see \textbf{Sec.~\ref{sec:Various Retrievers}}).
We further incorporate a security layer that combines rewriting strategies and guardrail models (\textbf{Sec.\ref{Sec:Robustness Against Defense Mechanisms}}).
Specifically, the rewriting strategies include query rewriting and response rewriting~\cite{ma2023query,dharwada2025queryrewritingllms,wang2025maferw}, while the guardrail models include PI-Guard$\footnote{PI-Guard: https://huggingface.co/leolee99/PIGuard}$~\cite{li2025piguard}, Tanaos-Guardrail$\footnote{Tanaos-Guardrail-v1: https://huggingface.co/tanaos/tanaos-guardrail-v1}$, and FT-Llama-Prompt-Guard-2$\footnote{FT-Llama-Prompt-Guard-2: https://huggingface.co/Aira-security/FT-Llama-Prompt-Guard-2}$~\cite{ft_llama_prompt_guard_2}.
Together, these components monitor and sanitize the interaction pipeline by detecting malicious inputs and filtering unsafe outputs, ensuring robust and consistent RAG behavior during evaluation. Unless specified, we use Top $k$=3, retriever = BGE~\cite{zhang2023retrieve}, exam question numbers N = 28 as default.

\par\noindent
{\bf Baselines and Evaluation Metrics.}
We compare \alg\ with three RAG-MIA baselines (S$^2$MIA~\cite{li2025generating}, MBA~\cite{liu2025mask}, and IA~\cite{naseh2025riddle}). 
\begin{itemize}[leftmargin=*]
    \item \textbf{S$^2$MIA}~\cite{li2025generating} infers membership by measuring semantic similarity (\emph{e.g.}, BLEU) between RAG outputs and target documents, assuming higher overlap for in-corpus data.
    \item \textbf{MBA}~\cite{liu2025mask} estimates membership via masked token completion, using the model’s accuracy in reconstructing hidden content as a signal of familiarity.
    \item \textbf{IA}~\cite{naseh2025riddle} formulates membership inference as a series of Yes/No questions derived from document summaries, and evaluates response accuracy to determine membership.
\end{itemize}

We report Accuracy (Acc), AUC-ROC, AUC-PR, and TPR@FPR (0.05/0.01/0.005) as evaluation metrics. 
In our evaluation setup, the number of member and non-member samples is balanced, allowing this metric to reflect the model’s predictive performance on both classes. 

\input{B-5-Experiments-RQ1}

\input{B-5-Experiments-RQ2}

\input{B-5-Experiments-RQ3}

\input{B-5-Experiments-RQ4}

\input{B-5-Experiments-RQ5}

%% file: B-5-Experiments-RQ1.tex
\subsection{Overall Performance (RQ1)}
\label{sec:performance comparison}
Table~\ref{tab:overall} compares our proposed E-MIA with three baselines across three RAG corpora and four LLM backbones.
\alg\ demonstrates substantial performance gains over baselines across all datasets and models. 
It consistently achieves near-perfect scores, with Accuracy exceeding 0.97 and AUC-ROC often reaching 1.000.
The most striking advantage of E-MIA is its stability under high-precision constraints. 
At an extremely low false positive rate (FPR=0.005), baseline methods often suffer a catastrophic performance collapse, with their True Positive Rates (TPR) frequently plummeting to values falling below 0.40.
In sharp contrast, E-MIA sustains robust discriminative power, often exceeding 0.980 and reaching as high as 0.998 in many configurations.
This resilience is reflected in the average improvement score, where E-MIA improved TPR@FPR=0.005 by an astonishing margin, with a minimum improvement of 117.8\% and a maximum improvement of 142.6\% relative to the best-performing baseline. 
These results demonstrate that \alg\ fundamentally amplifies the distinctive signal evidence embedded in documents, evidence that conventional black-box MIA methodologies consistently fail to capture.

%% file: B-5-Experiments-RQ2.tex
\subsection{Mechanism Analysis (RQ2)}
To better understand why E-MIA achieves superior performance, we analyze its underlying mechanisms from multiple perspectives. 
Specifically, we first examine the effectiveness of the structural design, then investigate how different types of evidence are mapped to question formats, and finally analyze the resulting score distributions to reveal the separability between member and non-member samples.

\par
\begin{figure}[!h]
    \centering
    \subfigure[Different blank counts in FB]{
    \includegraphics[width=0.45\linewidth]{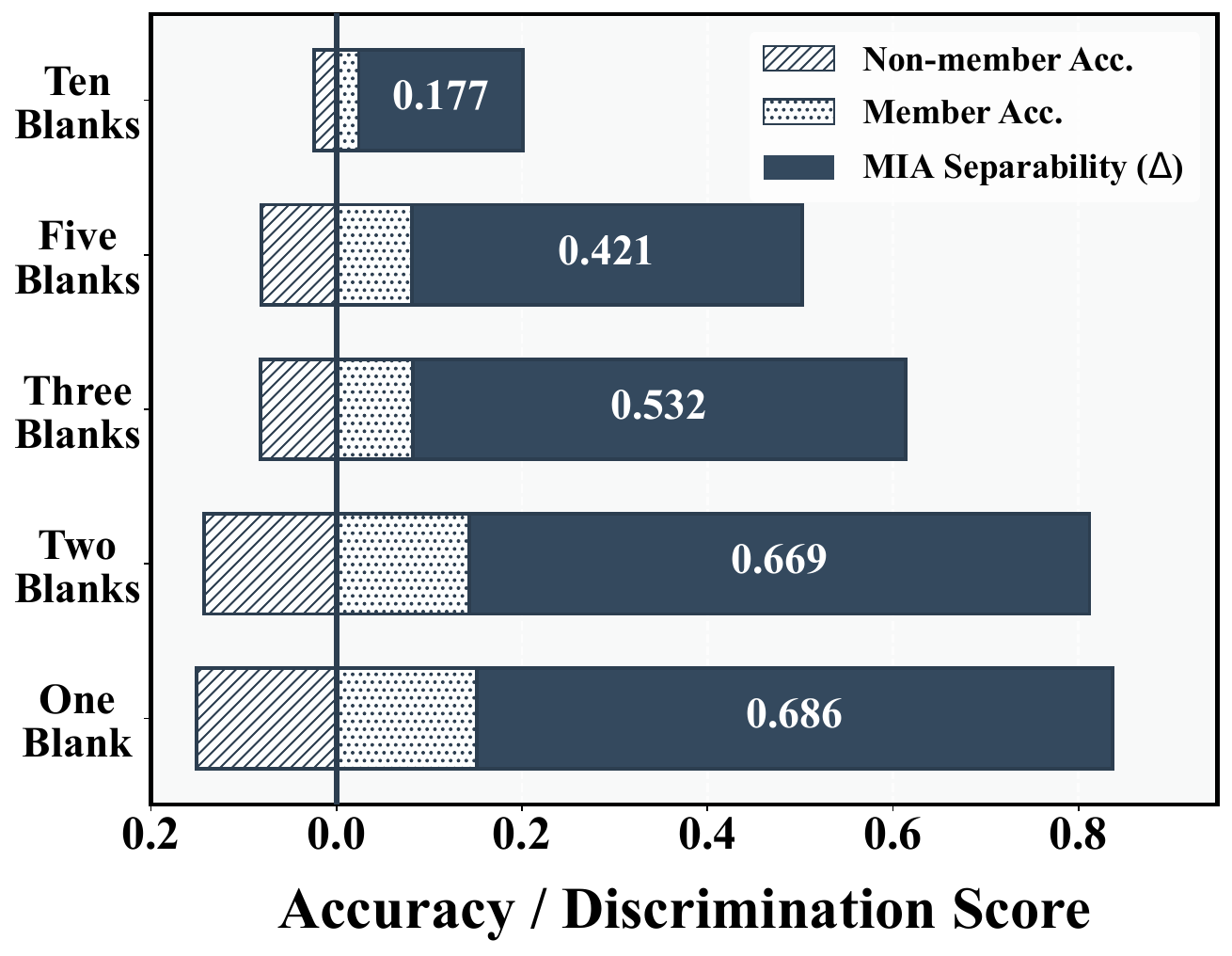}
    }
    \vspace{-0.7em}
    \subfigure[Option Design Factors in MC]{
    \includegraphics[width=0.45\linewidth]{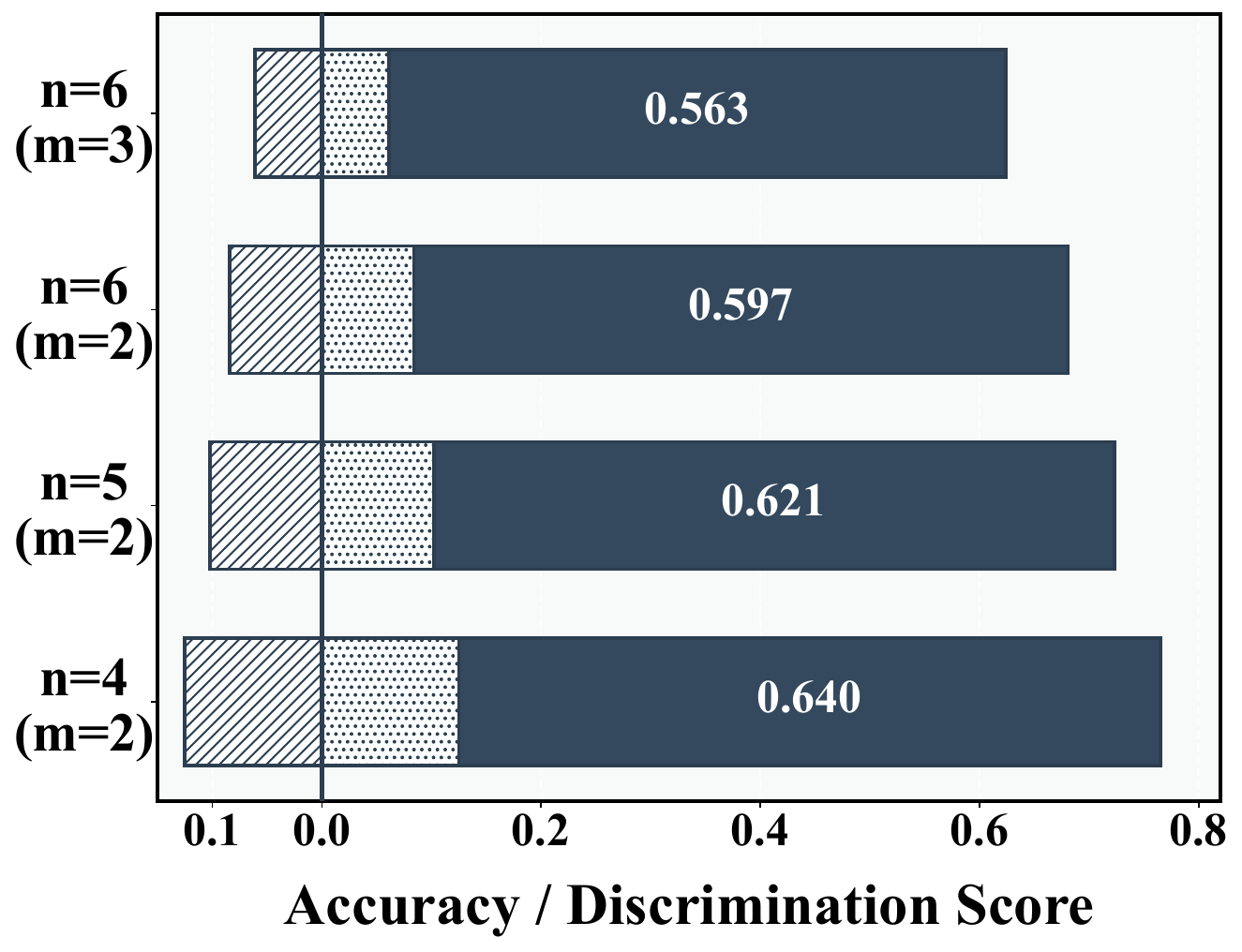}
    }
    \\
    \subfigure[Option Design Factors in SC]{
    \includegraphics[width=0.45\linewidth]{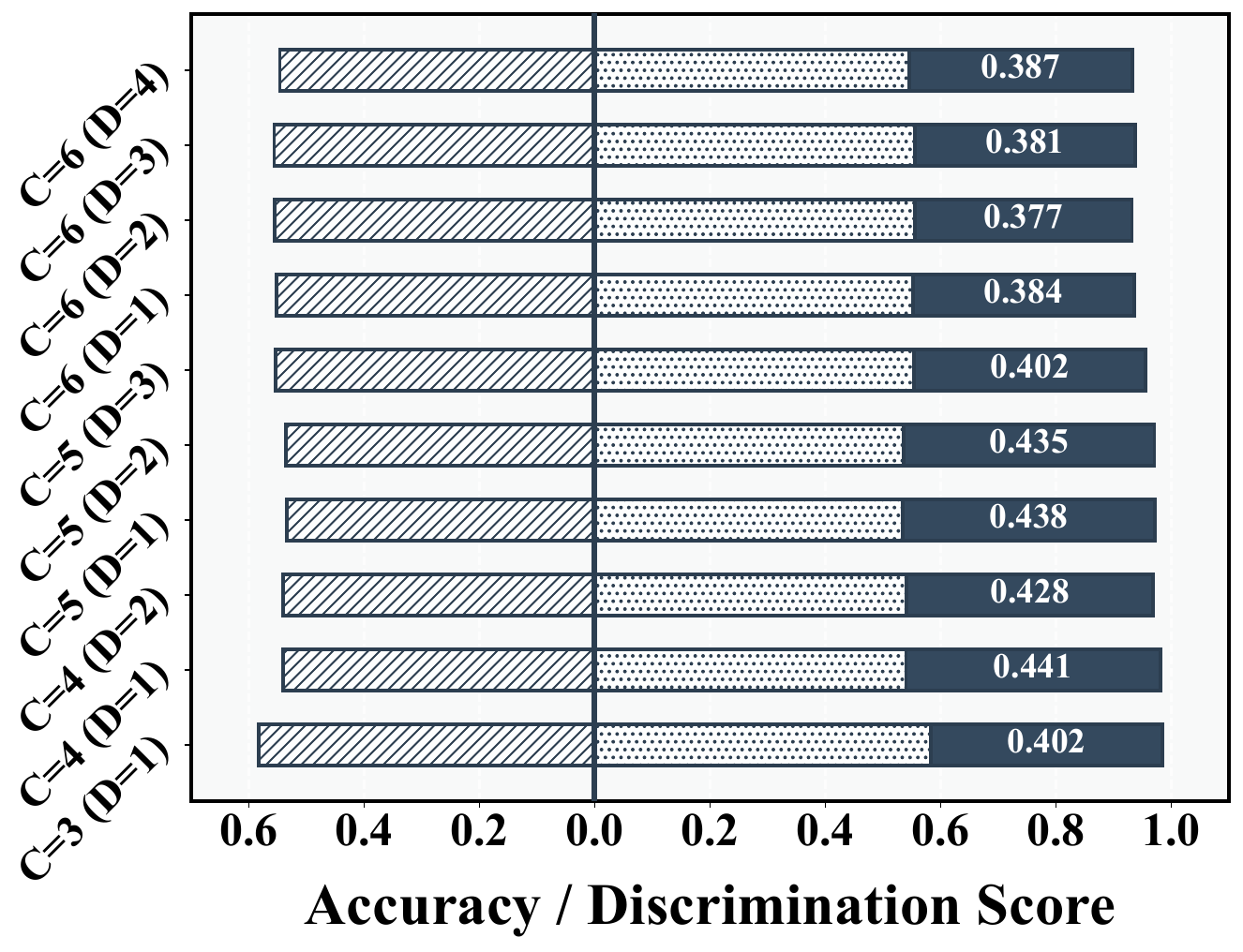}
    }
    \subfigure[Question type combination]{
    \includegraphics[width=0.45\linewidth]{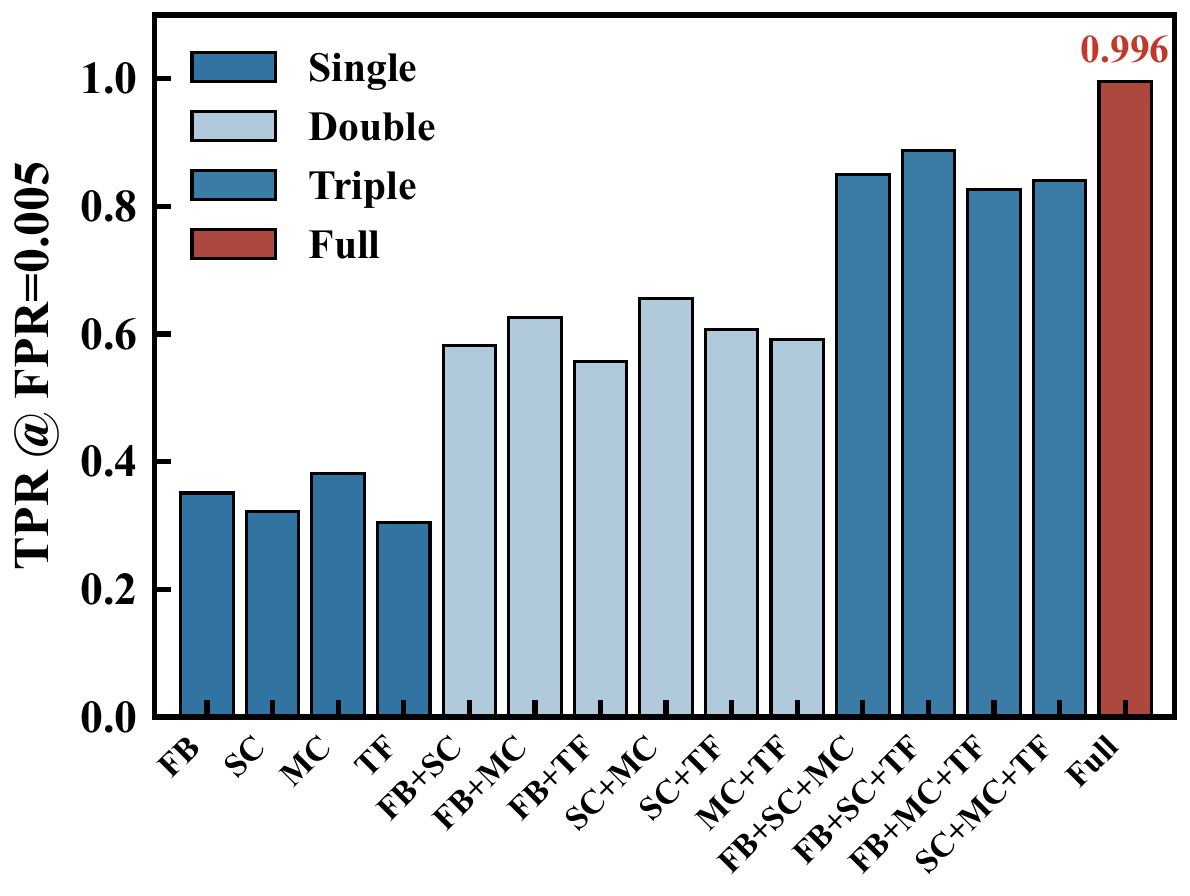}}
    \caption{Exploring the effectiveness of question types}
    \label{fig:Question}
\end{figure}

\noindent
\textbf{Structural Effectiveness}.
\label{ablation and effectiveness studies}
To maximize discriminative power, we conducted a granular distribution analysis on individual question parameters, as illustrated in Fig.~\ref{fig:Question}(a)-(c), focusing on membership separability ($\Delta$) to identify robust configurations. 
For SC, we found that $C \in \{4,5\}$ options with $D \in \{1,2\}$ similar distractors yields stable, high discrimination. 
For FB, one or two blanks are optimal, excessive blanks increase difficulty such that even models with the retrieved target document fail to answer correctly. 
For MC, we adopt $n \in \{4,5\}$ options with $m=2$ correct answers to ensure peak separability.
Beyond question design, Fig.~\ref{fig:Question}(d) confirms that format diversity significantly amplifies the membership signal. 
Integrating all question types into a "Full (All)" configuration achieves a peak TPR@FPR=0.005 of 0.996, consistently outperforming any single-format approach. 
These results demonstrate that \alg's heterogeneous exam structure effectively leverages the multi-faceted nature of model memorization to maximize inference effectiveness. 

\par
\begin{figure}[h!]
     \centering
     \includegraphics[width=0.9\linewidth]{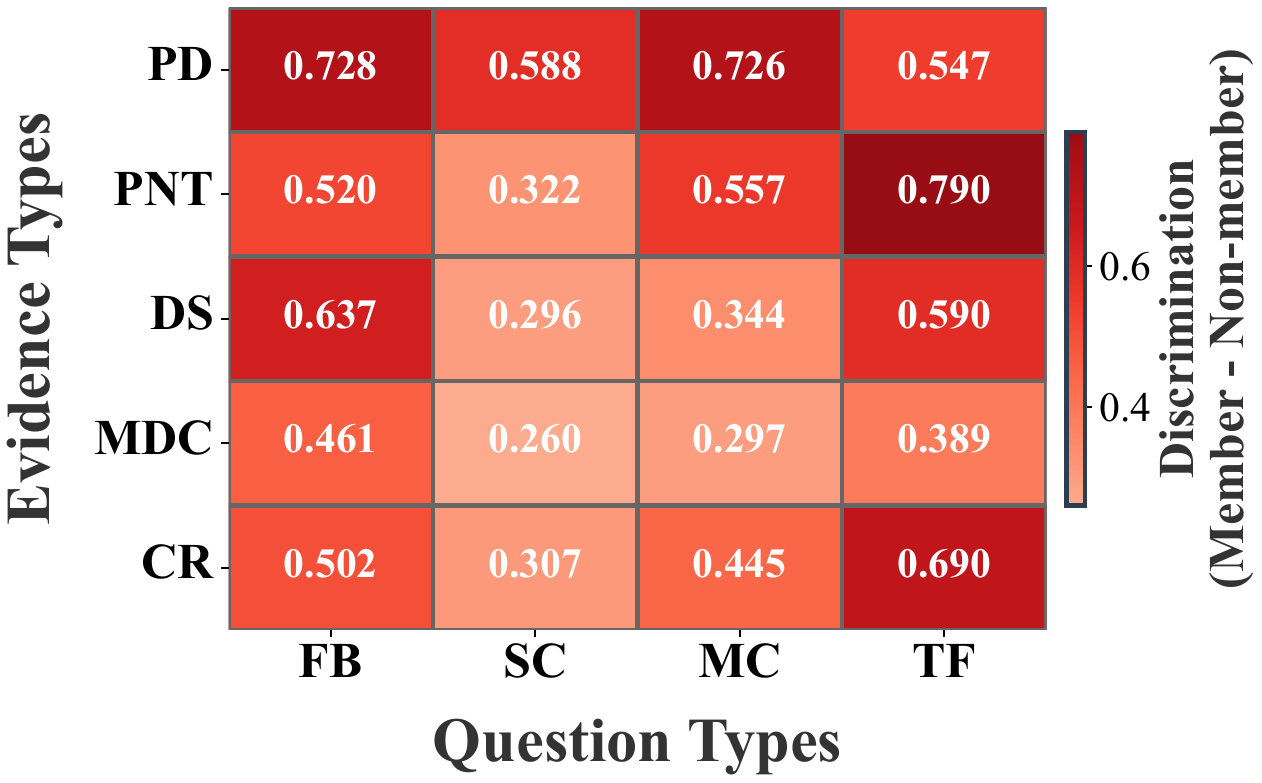}
     \caption{Heatmap of membership separability ($\Delta$) across evidence and question types}
     \label{fig:evidence}
\end{figure}

\noindent
\textbf{Impact of evidence-question matching}.
\label{Sec:matching}
To optimize membership signal extraction, we establish a bidirectional priority for evidence-to-question mapping based on the discriminative performance observed in Fig.~\ref{fig:evidence}. 
When a specific type of evidence is extracted, we prioritize assigning it to its most effective question format: Proper names and technical terms (PNT) and Constraint relations (CR) are predominantly mapped to TF questions to exploit the model's high-confidence binary certainty regarding these specific entities and logical links. 
Meanwhile, Precise details (PD) and Definitional statements (DS) are funneled into FB tasks to demand the high-fidelity textual reproduction of numerical anchors and concept boundaries. 
Metadata cues (MDC), such as structural markers, are assigned to SC questions to filter out common knowledge via semantically similar distractors. 
This evidence-driven prioritization ensures that each factual "anchor" is queried using the format that maximizes membership separability ($\Delta$).

Conversely, the selection of evidence for each question type follows a hierarchical preference to maintain robust detection results. 
For TF questions, we prioritize PNT and CR as primary sources, as these yield the sharpest contrast between indexed and unseen documents, reaching a peak $\Delta$ of 0.790. 
For FB questions, the priority shifts toward PD and DS, as numbers and exact definitions provide more stable "hard signals" than metadata. 
MC questions are prioritized for multi-attribute verification, particularly utilizing PD to aggregate several facts into a single complex item. 
This dual-optimization strategy ensures that the structural design of \alg\ successfully exploits multi-dimensional memorization features, converting unique document fingerprints into a potent and discriminative membership signal.

\par
\begin{figure}[!h]
    \centering
    \vspace{-1.5em}
    \subfigure[E-MIA(ours)]{
    \includegraphics[width=0.45\linewidth]{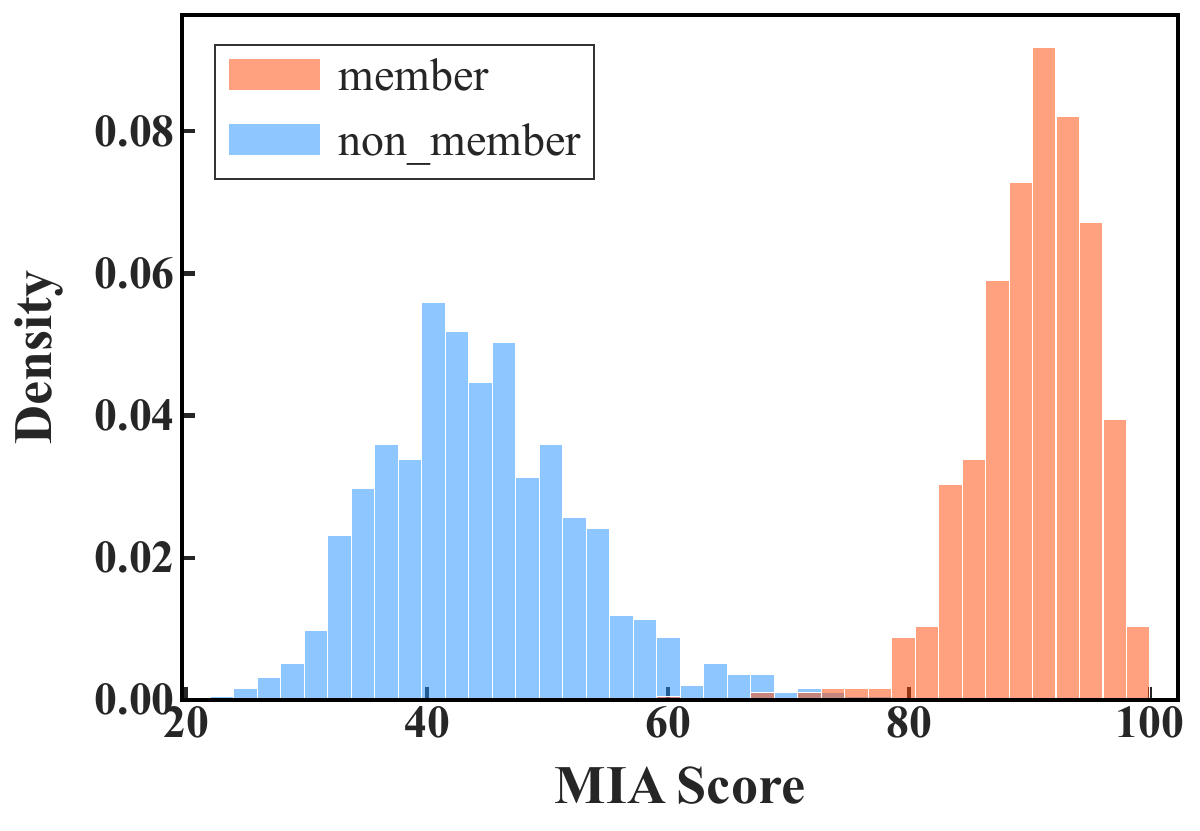}
    \label{fig:main-E-MIA}
    }
    \vspace{-0.7em}
    \subfigure[S$^2$MIA]{
    \includegraphics[width=0.45\linewidth]{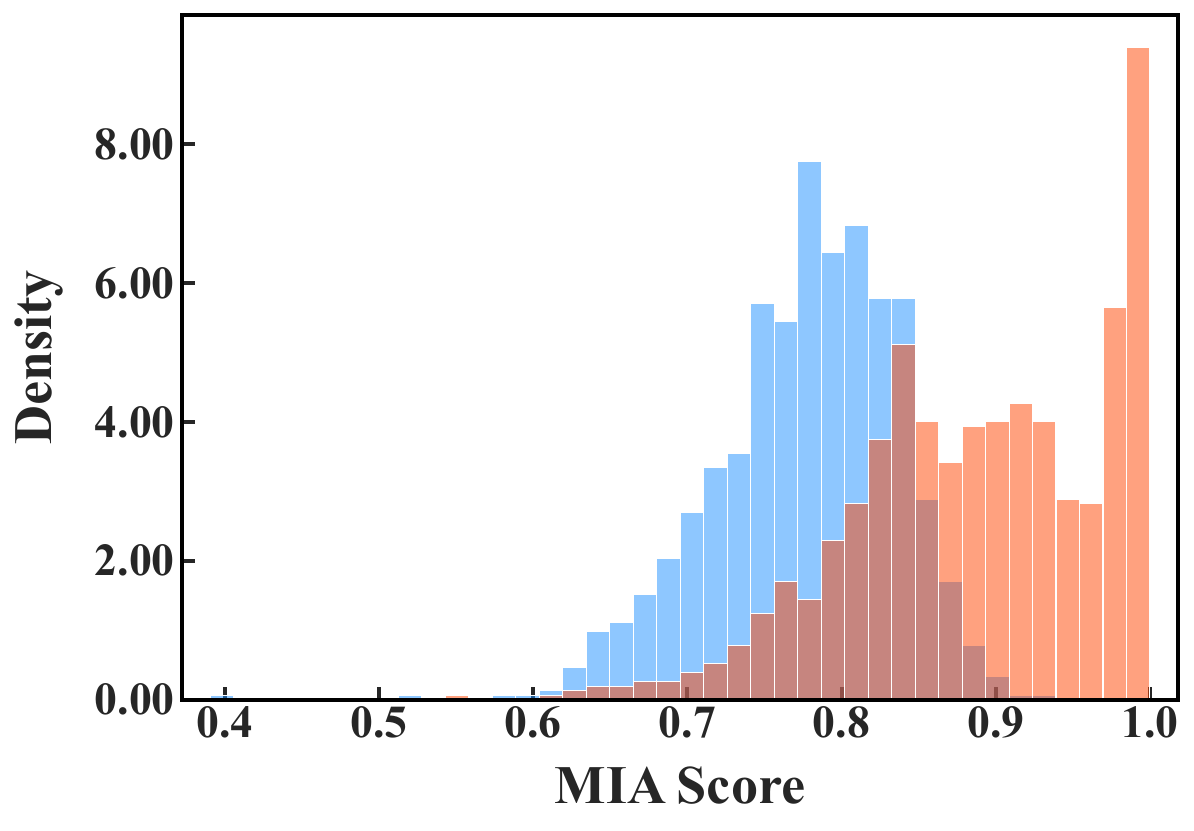}
    \label{fig:main-S$^2$MIA}
    }
    \\
    \subfigure[MBA]{
    \includegraphics[width=0.45\linewidth]{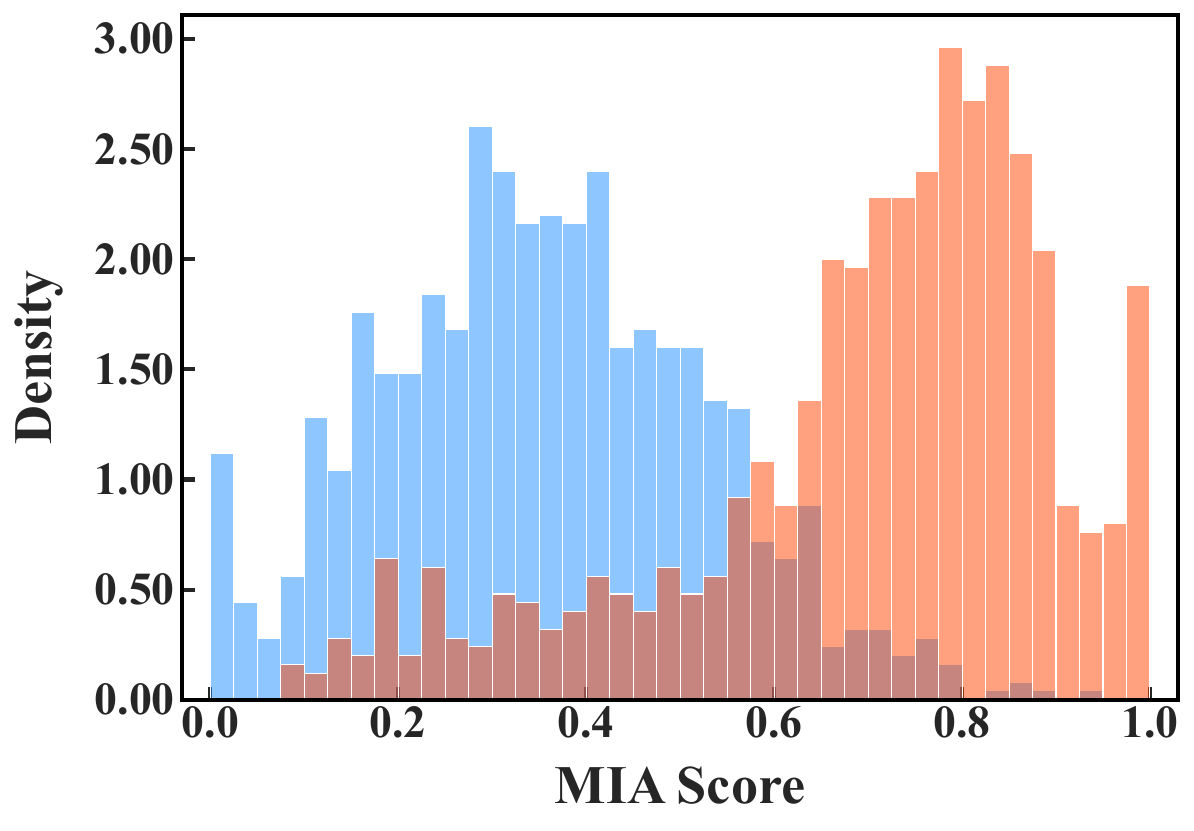}
    \label{fig:main-MBA}
    }
    \subfigure[IA]{
    \includegraphics[width=0.45\linewidth]{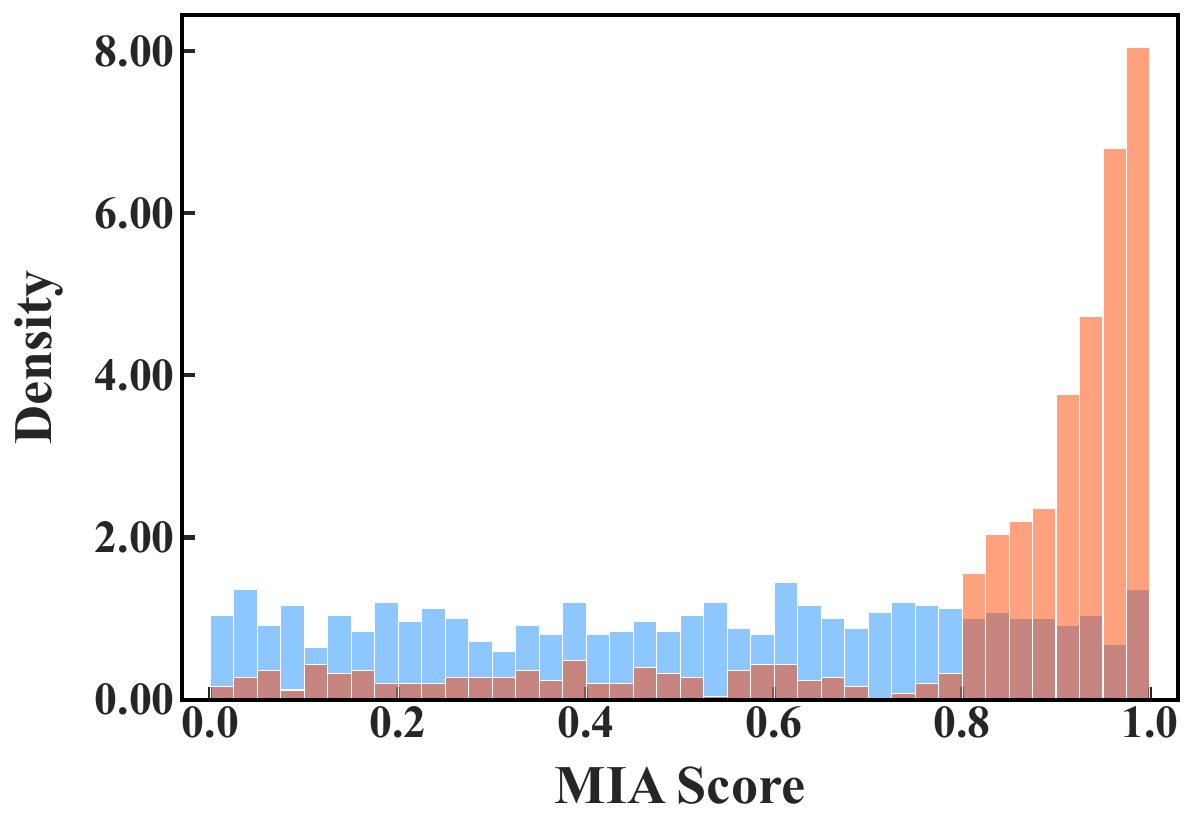}
    \label{fig:main-IA}
    }
    \caption{Membership and non-membership score distribution}
    \label{fig:scoredistribution-main}
\end{figure}

\noindent
\textbf{Distributional Advantage.} 
\label{sec:Distributional Advantage}
We visualize the membership score distributions for E-MIA and three competitive baselines on the TREC-COVID dataset in Fig.~\ref{fig:scoredistribution-main}. 
As illustrated in the density plots, \alg\ (Fig. \ref{fig:main-E-MIA}) achieves a significantly clearer and wider bi-modal separation between the member (red) and non-member (blue) groups compared to its counterparts. 
While baseline methods like S$^2$MIA (Fig. \ref{fig:main-S$^2$MIA}), MBA (Fig. \ref{fig:main-MBA}), and IA (Fig. \ref{fig:main-IA}) exhibit substantial overlapping regions, indicating a higher likelihood of false positives and a weaker membership signal, \alg\ effectively pushes the two distributions toward opposite ends of the score spectrum.

Specifically, the minimal intersection in \alg's distribution confirms that it amplifies the intrinsic signal gap between indexed knowledge base documents and unseen documents.  
By minimizing the density in the overlapping region, \alg\ effectively isolates document-level memorization signatures, ensuring that high-confidence membership inferences are directly anchored to verifiable document evidence rather than ambiguous similarity patterns. 
This superior discriminative capability stems from \alg's document-centric design, which transforms the RAG generator's response into a structured examination of document-specific evidence, thereby capturing the multi-dimensional memorization features that remain obscured in traditional one-dimensional similarity metrics.

%% file: B-5-Experiments-RQ3.tex
\subsection{Robustness Analysis (RQ3)}
To comprehensively evaluate the robustness of \alg, we conduct experiments under multiple challenging conditions. 
We first investigate its resilience against a range of RAG security defenses, and then analyze how varying the retrieval budget affects its stability and performance. 
These analyses provide insights into the reliability of E-MIA in practical deployment scenarios.

\par\noindent
\textbf{Robustness Against Defense Mechanisms}.
\label{Sec:Robustness Against Defense Mechanisms}
We further evaluated the resilience of \alg\ against common RAG security measures (Retrieval recall is analysis-only and not attacker-observable). 
As demonstrated in Table~\ref{tab:Rewriting}, \alg\ maintains exceptional performance under various rewriting defenses, sustaining near-perfect Accuracy (0.988) even under the most challenging "Both Rewriting" scenario with a negligible degradation of only 0.002, while baseline methods exhibit significant vulnerability. 
Beyond defense resilience, Table~\ref{tab:Detection} confirms the superior stealthiness of our approach, by framing membership queries as legitimate educational "exams", \alg\ consistently achieves near-perfect pass rates (up to 1.000) across representative security guardrails like PIGuard and Prompt-Guard.
This proves that \alg\ effectively circumvents modern security filters by utilizing benign-appearing inputs that traditional similarity-based or adversarial detection models fail to block. 
The high pass rates across diverse guardrails indicate that our structural transformation effectively decouples membership signals from the adversarial patterns typically flagged by safety classifiers. 
Unlike traditional attacks that rely on detectable semantic overlaps, \alg\ exploits the model's fundamental reasoning behavior within a benign pedagogical context, making it indistinguishable from standard user interactions. 

\begin{table}[!t]
    \centering
    \caption{Robustness of MIAs against rewriting strategies}
    \label{tab:Rewriting}
    \renewcommand{\arraystretch}{0.95}
    \setlength{\tabcolsep}{6pt}
    \scalebox{0.9}{
        \begin{tabular}{c|c|cc}
        \toprule
        \midrule
        \textbf{Rewriting Strategy} & \textbf{Method} & \textbf{Retrieval Recall} & \textbf{Accuracy} \\ 
        \midrule
        \multirow{4}{*}[0pt]{\textbf{Query Rewriting}} & S$^2$MIA & 0.942(0.041$\downarrow$) & 0.722(0.019$\downarrow$) \\
        & MBA & 0.971(0.011$\downarrow$) & 0.597(0.010$\downarrow$) \\
        & IA & 0.871(0.115$\downarrow$) & 0.801(0.181$\downarrow$) \\
        & E-MIA(Ours) & \raisebox{-1pt}{\textbf{\small 1.000(0.000$\downarrow$)}} & \raisebox{-1pt}{\textbf{\small 0.988(0.002$\downarrow$)}} \\
        \cmidrule(lr){1-4}
        \multirow{4}{*}[0pt]{\textbf{Response Rewriting}} & S$^2$MIA & 0.982(0.001$\downarrow$) & 0.484(0.257$\downarrow$) \\
        & MBA & 0.979(0.003$\downarrow$) & 0.601(0.006$\downarrow$) \\
        & IA & 0.986(0.000$\downarrow$) & 0.963(0.019$\downarrow$) \\
        & E-MIA(Ours) & \raisebox{-1pt}{\textbf{\small 1.000(0.000$\downarrow$)}} & \raisebox{-1pt}{\textbf{\small 0.987(0.003$\downarrow$)}} \\
        \cmidrule(lr){1-4}
        \multirow{4}{*}[0pt]{\textbf{Both Rewriting}} & S$^2$MIA & 0.950(0.033$\downarrow$) & 0.462(0.279$\downarrow$) \\
        & MBA & 0.972(0.010$\downarrow$) & 0.598(0.009$\downarrow$) \\
        & IA & 0.825(0.161$\downarrow$) & 0.691(0.291$\downarrow$) \\
        & E-MIA(Ours) & \raisebox{-1pt}{\textbf{\small 1.000(0.000$\downarrow$)}} & \raisebox{-1pt}{\textbf{\small 0.988(0.002$\downarrow$)}} \\
        \bottomrule
        \end{tabular}
    }
\end{table}

\begin{table}[!t]
    \centering
    \caption{Pass rates of various MIA methods across three different malicious input detection models}
    \setlength{\tabcolsep}{10pt}
    \renewcommand{\arraystretch}{0.85}
    \scalebox{0.9}{
    \label{tab:Detection}
    \begin{tabular}{c|c|c}
        \toprule
        \midrule
        \textbf{Detection Model} & \textbf{Method} & \textbf{Pass Rate} \\
        \midrule
        \multirow{4}{*}[0pt]{\textbf{PIGuard}} & S$^2$MIA & 0.439 \\
        & MBA & 0.533 \\
        & IA & 0.976 \\
        & E-MIA(Ours) & \raisebox{-1pt}{\textbf{\small 1.000}} \\
        \cmidrule(lr){1-3}
        \multirow{4}{*}[0pt]{\textbf{Tanaos-Guardrail-v1}} & S$^2$MIA & 0.984 \\
        & MBA & 0.653 \\
        & IA & 0.991 \\
        & E-MIA(Ours) & \raisebox{-1pt}{\textbf{\small 0.997}} \\
        \cmidrule(lr){1-3}
        \multirow{4}{*}[0pt]{\textbf{FT-Llama-Prompt-Guard-2}} & S$^2$MIA & 0.210 \\
        & MBA & 0.456 \\
        & IA & 0.992 \\
        & E-MIA(Ours) & \raisebox{-1pt}{\textbf{\small 1.000}} \\
        \bottomrule
    \end{tabular}
    }
\end{table}

%% file: B-5-Experiments-RQ4.tex
\subsection{Parameter Analysis (RQ4)}
\label{sec:Parameter Calibration}

\par
\begin{figure}[h!]
    \centering
    \subfigure[Optimal Parameters]{
    \includegraphics[width=0.46\linewidth]{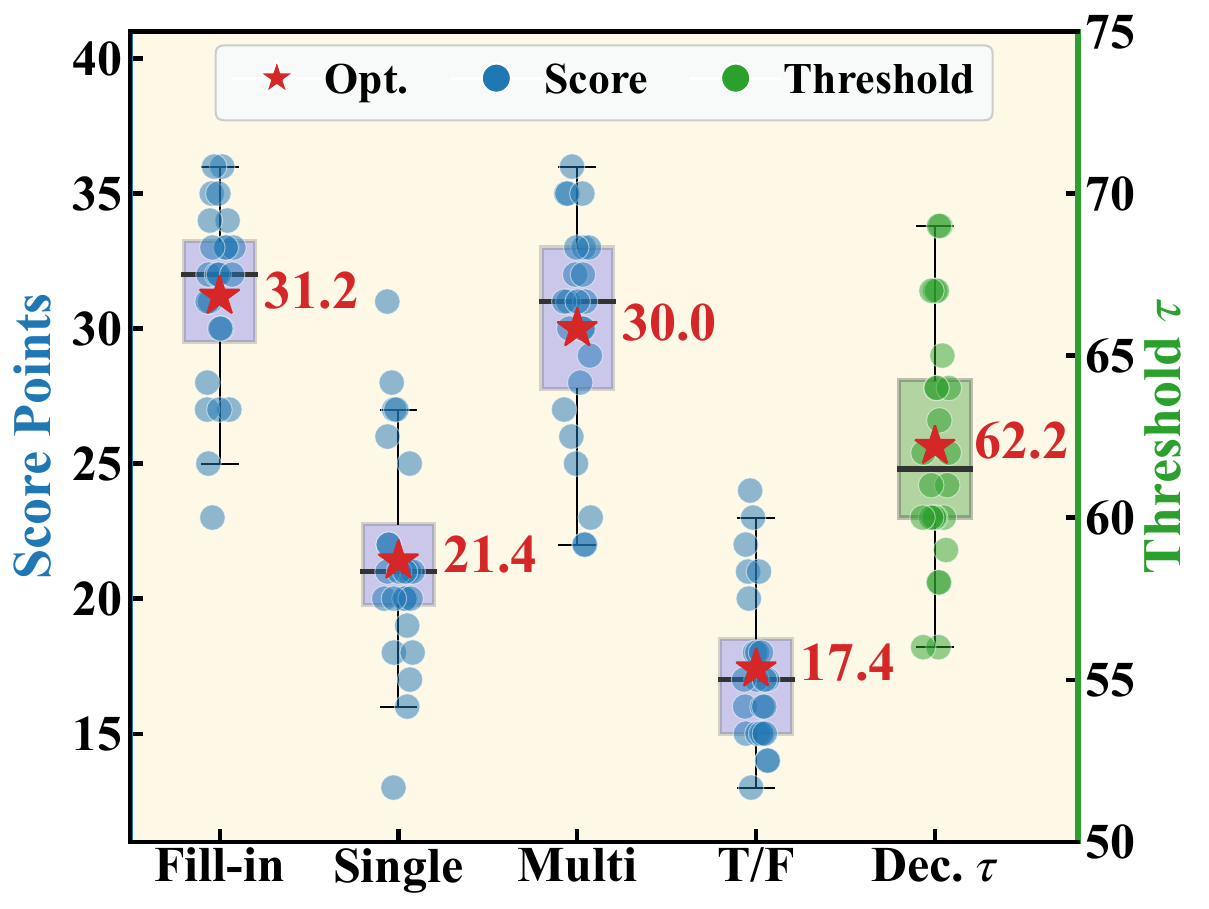}
    }
    \subfigure[Query Scalability]{
    \includegraphics[width=0.46\linewidth]{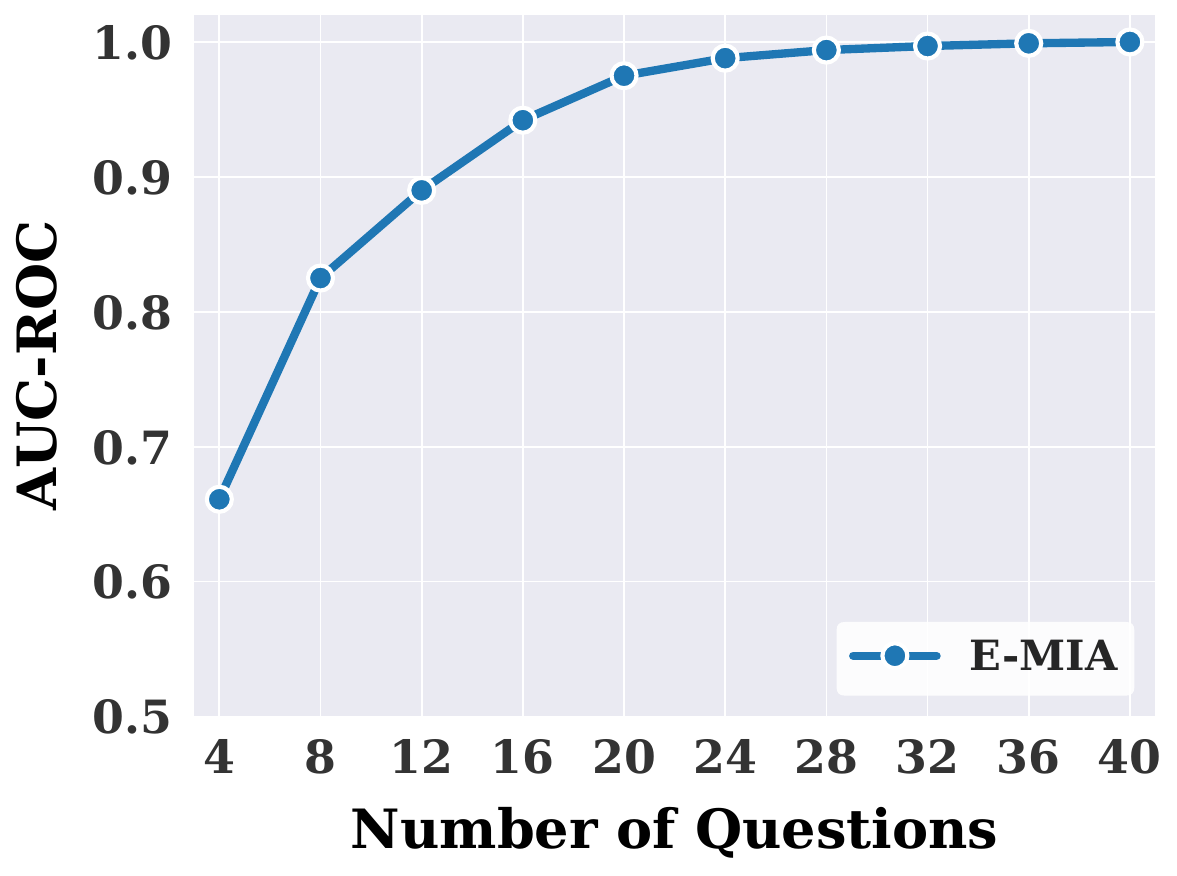}
    }
    \caption{Influence of Question Density and Design}
    \label{fig:44}
\end{figure}

\noindent
\textbf{Parameter Calibration}. 
To ensure the effectiveness of \alg\ in black-box scenarios, we conducted a systematic investigation into key parameters, including score allocation, the decision threshold $\tau$, and query volume, as illustrated in Fig.~\ref{fig:44}. 
Based on 25 independent trials conducted across diverse datasets and model architectures, we identified the Medoid configuration (marked by red stars in Fig.~\ref{fig:44}) as the most stable parameter set. 
Formally, a Medoid is defined as the specific observed data point within a cluster that minimizes the sum of dissimilarities to all other points in that set.
This selection process accounts for the inherent variance in model responses, ensuring that the chosen parameters generalize effectively even when the adversary has no prior knowledge of the target system's training distribution.

In addition to weighting, we evaluated the attack's scalability through an incremental query strategy. 
As shown in Fig.~\ref{fig:44}(b), the AUC-ROC follows a steep logarithmic ascent, achieving near-optimal performance with only 20 to 24 questions before plateauing. 
This rapid saturation proves that \alg\ consolidates membership signals with an extremely compact query budget.
Through this comprehensive analysis, we determined the finalized robust parameters: a score allocation of FB (31.2), SC (21.4), MC (30.0), and T/F (17.4), paired with an optimized decision threshold of $\tau = 62.2$. 
To ensure absolute reliability in practical deployments, we recommend a query volume of at least 28 questions (\emph{i.e.}, at least seven items per question type) to maximize attack efficacy with minimal overhead.

\par
\begin{figure}[h!]
     \centering
     \includegraphics[width=0.9\linewidth]{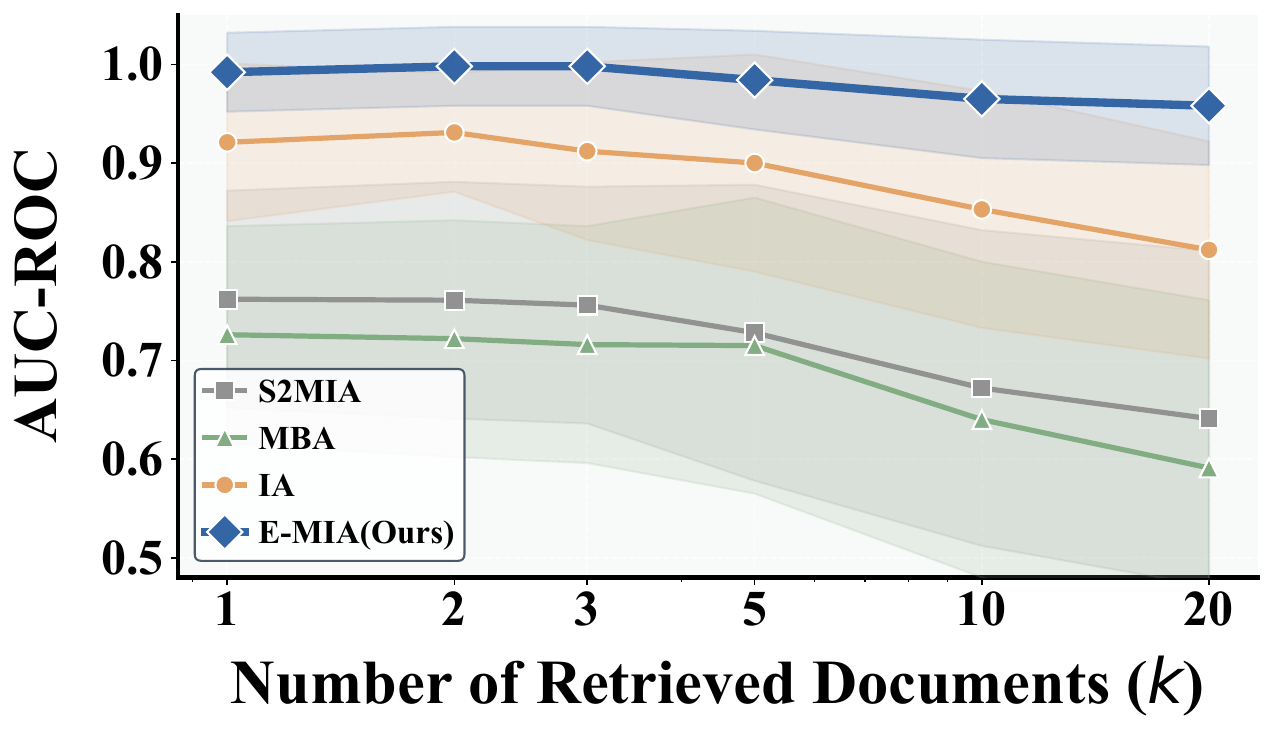}
     \caption{Performance stability across different numbers of retrieved documents ($k$)}
     \label{fig:Retrieved Documents}
\end{figure}

\noindent
\textbf{Effect of the Retrieval Budget ($k$)}. 
\label{Sec:Retrieval Budget}
We investigate the impact of retrieval scale on attack performance, evaluating E-MIA across varying numbers of retrieved documents ($k$), ranging from 1 to 20. 
As illustrated in Fig.~\ref{fig:Retrieved Documents}, our method demonstrates exceptional stability compared to competitive baselines. 
While methods such as S$^2$MIA, MBA, and IA exhibit a noticeable performance decay as $k$ increases, likely due to the introduction of noise from non-member documents, \alg\ maintains a consistently high AUC-ROC, staying near-optimal even at $k=20$.
This resilience stems from our structured exam design, which effectively isolates the non-member document's signal through specific question-answering tasks, making the inference process less sensitive to the surrounding context provided by the retriever. 
The results confirm that \alg\ is highly robust to the retrieval budget, ensuring reliable membership inference even in RAG systems configured with large retrieval windows.

%% file: B-5-Experiments-RQ5.tex
\subsection{Generalization Analysis (RQ5)}
To evaluate the generalization ability of E-MIA, we conduct experiments across different languages, generator models, and retrievers. 
We first assess its cross-lingual performance on a Chinese dataset, then evaluate its consistency across multiple generators, and finally analyze its robustness under different retrieval backends.

\par
\begin{figure}[t!]
     \centering
     \includegraphics[width=0.76\linewidth]{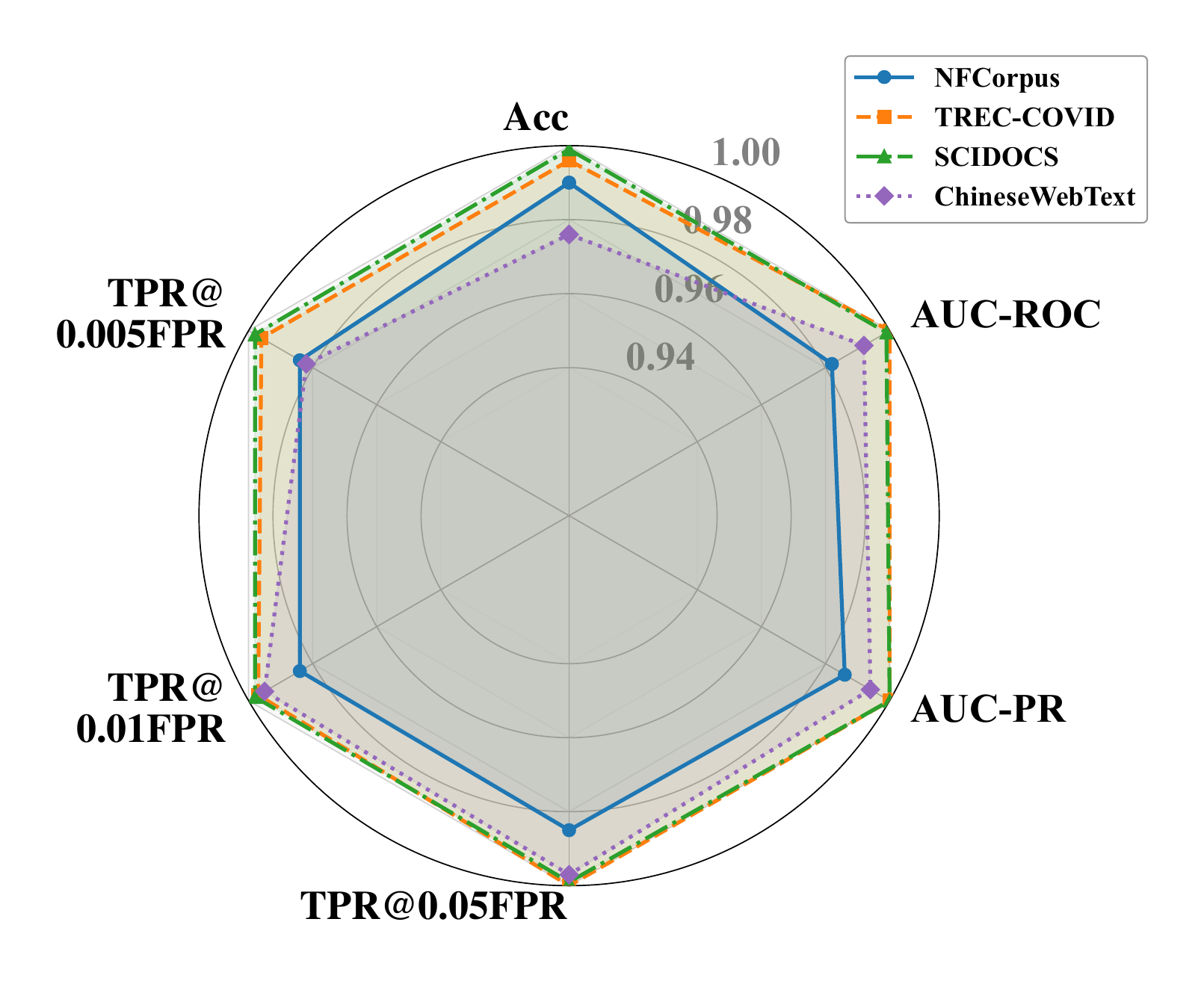}
     \caption{Performance across diverse language datasets}
     \label{fig:chinese}
\end{figure}

\noindent
\textbf{Performance on Chinese Datasets}.
\label{Sec:generators}
We also assess the cross-lingual generalizability of \alg\ by extending our evaluation to the Chinese dataset, ChineseWebText~\cite{chen2023chinesewebtext}.
As illustrated in the radar chart in Fig.~\ref{fig:chinese}, \alg\ maintains its exceptional performance without any language-specific degradation, achieving near-perfect scores across all key metrics (\emph{i.e., Accuracy, AUC-ROC, and TPR at various FPR levels}), regardless of the linguistic context.

The high degree of consistency between the English and Chinese results confirms that the membership signals captured by our multi-dimensional exam structure are rooted in the intrinsic memorization patterns of large language models rather than specific linguistic features. 
This stability across languages suggests that the structured exam format effectively bypasses surface-level syntax to probe the underlying semantic retention of the model, which remains uniform across different language. 
Furthermore, the ability to achieve near-optimal TPR even at the most stringent FPR levels ($0.005$) on both Chinese and English sets demonstrates that~\alg\ is a robust, language-agnostic framework capable of reliably identifying training data membership in globalized RAG applications. 
By effectively normalizing the score distribution across different languages, our method mitigates the risk of language-specific bias in privacy auditing, providing a standardized tool for cross-lingual data protection assessments.

\par
\begin{figure}[!h]
     \centering
     \includegraphics[width=0.9\linewidth]{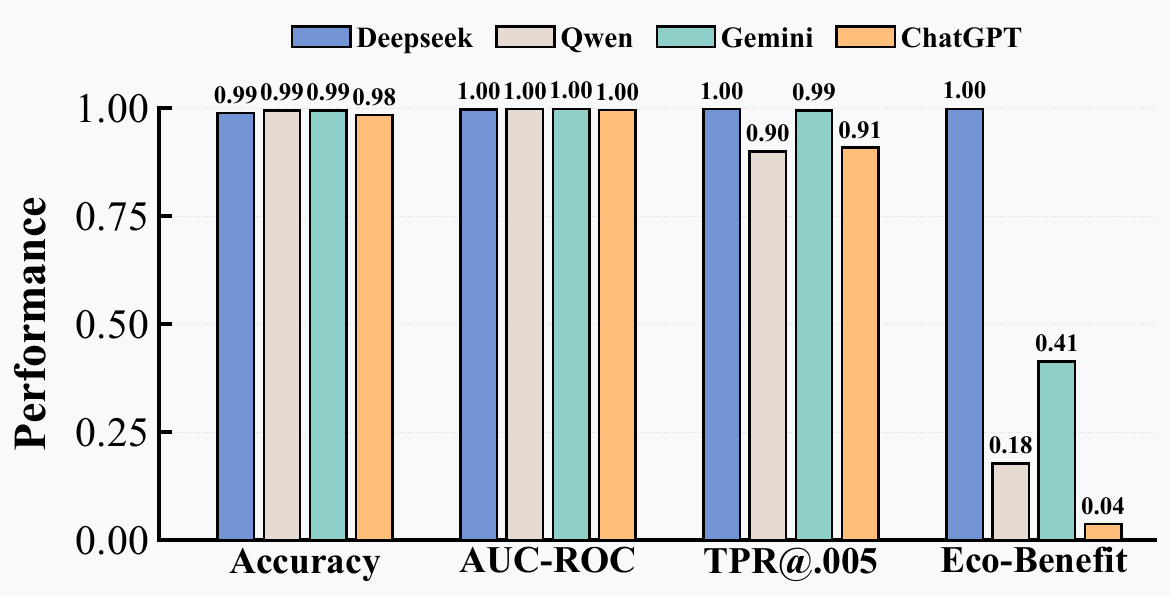}
     \caption{Performance Across Diverse Exam Generators}
     \label{fig:performance_comparison}
\end{figure}

\noindent
\textbf{Effect of the Exam Generator Model}.
\label{Sec:generators}
To verify that E-MIA's effectiveness is not dependent on a specific generative architecture, we extended our evaluation to include three additional commercial APIs: Gemini-2.0-Flash~\cite{comanici2025gemini}, Qwen-Max~\cite{yang2025qwen3}, and GPT-5-mini~\cite{singh2025openai}.
As illustrated in Fig.~\ref{fig:performance_comparison}, \alg\ demonstrates remarkable consistency across all tested generators. 
Accuracy values cluster tightly between 0.985 and 0.995, while AUC-ROC scores consistently reach near-perfection, ranging from 0.996 to 1.000. 
Even under the most stringent precision constraints, the TPR@FPR=0.005 metric remains above 0.900 for all models.
But we designate DeepSeek-R1 as our primary exam generator due to its superior balance of high-fidelity performance and economic feasibility. 
As shown by the Eco-Benefit metric in Fig.~\ref{fig:performance_comparison}, we normalize cost as $\text{Eco}=\text{Cost}_{\min}/\text{Cost}_{\text{model}}$, where $\text{Cost}_{\min}$ is the minimum average cost per exam (thus the cheapest model, DeepSeek-R1, has $\text{Eco}=1$). 
DeepSeek-R1 incurs the lowest overhead among the evaluated models, with an average cost of only $\$0.0063$ per target document (\emph{i.e.}, per generated exam).

\par
\begin{figure}[!h]
  \centering
  \includegraphics[width=0.45\textwidth]{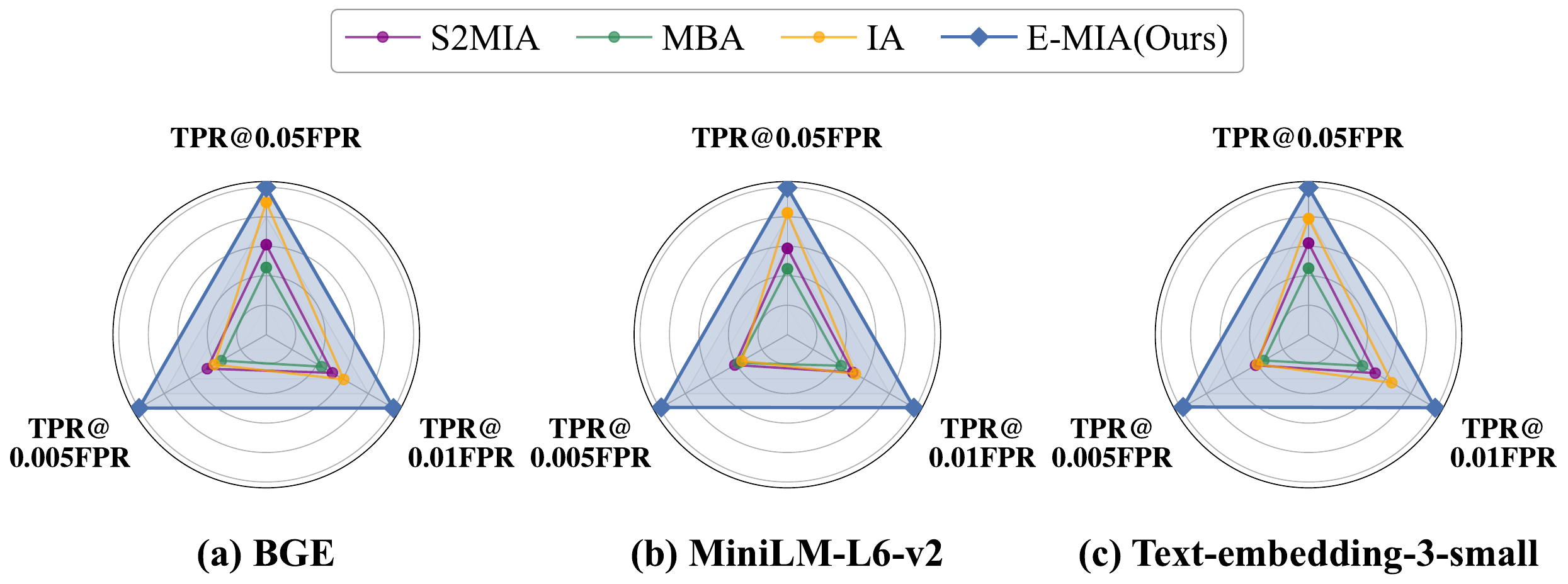}
  \caption{Performance of E-MIA across various retrievers}
  \label{fig:retrievers}
\end{figure}

\noindent
\textbf{Various Retrievers}.
\label{sec:Various Retrievers}
To evaluate whether the effectiveness of E-MIA is sensitive to the underlying retrieval component, we conducted a robustness analysis across three widely used embedding models: BGE, MiniLM-L6-v2, and Text-embedding-3-small. 
As illustrated in the radar charts in Fig. \ref{fig:retrievers}, \alg\ demonstrates remarkable stability regardless of the specific retriever employed.
While baseline methods such as S$^2$MIA, MBA, and IA show significant performance fluctuations depending on the retriever's embedding quality, \alg\ consistently maintains near-optimal scores across all key metrics, including TPR at various FPR levels (0.05, 0.01, and 0.005). 
This sustained performance suggests that our multi-dimensional exam structure successfully isolates membership signals that are independent of the initial document ranking or embedding space. 
The consistency observed across these diverse retrieval architectures validates \alg\ as a robust framework, ensuring reliable privacy auditing even in RAG systems with varying retrieval capabilities.

%% file: B-7-Conclusion.tex
\section{Conclusion}
\label{sec:conclusion}
In this work, we reveal a practical membership leakage channel in black-box RAG systems: document-specific \emph{hard evidence} can be reliably elicited through natural-looking QA prompts even when retrieval traces are hidden. 
To exploit this signal, we propose \alg, an exam-style document-level MIA framework that decomposes a target document into verifiable evidence, instantiates it into four objectively-gradable question types (FB/SC/MC/T/F), and aggregates black-box outcomes into a calibrated inference score for membership decisions. 
Extensive experiments across multiple datasets, retrievers, generators, and defense pipelines demonstrate that \alg\ achieves strong and stable separability, particularly under stringent low-FPR regimes, and remains robust to practical factors such as retrieval budget and the choice of exam generator. 
These results highlight that RAG deployments may expose privacy risks beyond coarse similarity cues, underscoring the need for stronger mitigations against evidence-level leakage. 